
\magnification=\magstephalf
\newread\epsffilein    
\newif\ifepsffileok    
\newif\ifepsfbbfound   
\newif\ifepsfverbose   
\newdimen\epsfxsize    
\newdimen\epsfysize    
\newdimen\epsftsize    
\newdimen\epsfrsize    
\newdimen\epsftmp      
\newdimen\pspoints     
\pspoints=1bp          
\epsfxsize=0pt         
\epsfysize=0pt         
\def\epsfbox#1{\global\def\epsfllx{72}\global\def\epsflly{72}%
   \global\def\epsfurx{540}\global\def\epsfury{720}%
   \def\lbracket{[}\def\testit{#1}\ifx\testit\lbracket
   \let\next=\epsfgetlitbb\else\let\next=\epsfnormal\fi\next{#1}}%
\def\epsfgetlitbb#1#2 #3 #4 #5]#6{\epsfgrab #2 #3 #4 #5 .\\%
   \epsfsetgraph{#6}}%
\def\epsfnormal#1{\epsfgetbb{#1}\epsfsetgraph{#1}}%
\def\epsfgetbb#1{%
%
%
\openin\epsffilein=#1
\ifeof\epsffilein\errmessage{I couldn't open #1, will ignore it}\else
%
%
   {\epsffileoktrue \chardef\other=12
    \def\do##1{\catcode`##1=\other}\dospecials \catcode`\ =10
    \loop
       \read\epsffilein to \epsffileline
       \ifeof\epsffilein\epsffileokfalse\else
%
%
          \expandafter\epsfaux\epsffileline:. \\%
       \fi
   \ifepsffileok\repeat
   \ifepsfbbfound\else
    \ifepsfverbose\message{No bounding box comment in #1; using defaults}\fi\fi
   }\closein\epsffilein\fi}%
%
%
\def\epsfsetgraph#1{%
   \epsfrsize=\epsfury\pspoints
   \advance\epsfrsize by-\epsflly\pspoints
   \epsftsize=\epsfurx\pspoints
   \advance\epsftsize by-\epsfllx\pspoints
%
%
   \epsfxsize\epsfsize\epsftsize\epsfrsize
   \ifnum\epsfxsize=0 \ifnum\epsfysize=0
      \epsfxsize=\epsftsize \epsfysize=\epsfrsize
%
%
     \else\epsftmp=\epsftsize \divide\epsftmp\epsfrsize
       \epsfxsize=\epsfysize \multiply\epsfxsize\epsftmp
       \multiply\epsftmp\epsfrsize \advance\epsftsize-\epsftmp
       \epsftmp=\epsfysize
       \loop \advance\epsftsize\epsftsize \divide\epsftmp 2
       \ifnum\epsftmp>0
          \ifnum\epsftsize<\epsfrsize\else
             \advance\epsftsize-\epsfrsize \advance\epsfxsize\epsftmp \fi
       \repeat
     \fi
   \else\epsftmp=\epsfrsize \divide\epsftmp\epsftsize
     \epsfysize=\epsfxsize \multiply\epsfysize\epsftmp
     \multiply\epsftmp\epsftsize \advance\epsfrsize-\epsftmp
     \epsftmp=\epsfxsize
     \loop \advance\epsfrsize\epsfrsize \divide\epsftmp 2
     \ifnum\epsftmp>0
        \ifnum\epsfrsize<\epsftsize\else
           \advance\epsfrsize-\epsftsize \advance\epsfysize\epsftmp \fi
     \repeat
   \fi
%
%
   \ifepsfverbose\message{#1: width=\the\epsfxsize, height=\the\epsfysize}\fi
   \epsftmp=10\epsfxsize \divide\epsftmp\pspoints
   \vbox to\epsfysize{\vfil\hbox to\epsfxsize{%
      \includegraphics{#1}%
      \hfil}}%
\epsfxsize=0pt\epsfysize=0pt}%

%
%
{\catcode`\%=12 \global\let\epsfpercent=
%
%
\long\def\epsfaux#1#2:#3\\{\ifx#1\epsfpercent
   \def\testit{#2}\ifx\testit\epsfbblit
      \epsfgrab #3 . . . \\%
      \epsffileokfalse
      \global\epsfbbfoundtrue
   \fi\else\ifx#1\par\else\epsffileokfalse\fi\fi}%
%
%
\def\epsfgrab #1 #2 #3 #4 #5\\{%
   \global\def\epsfllx{#1}\ifx\epsfllx\empty
      \epsfgrab #2 #3 #4 #5 .\\\else
   \global\def\epsflly{#2}%
   \global\def\epsfurx{#3}\global\def\epsfury{#4}\fi}%
%
%
\def\epsfsize#1#2{\epsfxsize}
%
%
\let\epsffile=\epsfbox
\def\Title#1{%
\vskip 1in{\titlefont\centerline{#1}}\vskip .5in}

\def\Date#1{\leftline{#1}\tenrm\supereject%
\global\hsize=\hsbody\global\hoffset=\hbodyoffset%
\footline={\hss\tenrm\folio\hss}}

\newif\ifdraftmode
\newif\ifleftlabels  

\def\nolabels{\def\wrlabeL##1{}\def\eqlabeL##1{}\def\reflabeL##1{}}
\def\writelabels{\def\wrlabeL##1{\leavevmode\vadjust{\rlap{\smash%
{\line{{\escapechar=` \hfill\rlap{\sevenrm\hskip.03in\string##1}}}}}}}%
\def\eqlabeL##1{{\escapechar-1\rlap{\sevenrm\hskip.05in\string##1}}}%
\def\reflabeL##1{\noexpand\rlap{\noexpand\sevenrm[\string##1]}}}
\def\writeleftlabels{\def\wrlabeL##1{\leavevmode\vadjust{\rlap{\smash%
{\line{{\escapechar=` \hfill\rlap{\sevenrm\hskip.03in\string##1}}}}}}}%
\def\eqlabeL##1{{\escapechar-1%
\rlap{\sixrm\hskip.05in\string##1}%
\llap{\sevenrm\string##1\hskip.03in\hbox to \hsize{}}}}%
\def\reflabeL##1{\noexpand\rlap{\noexpand\sevenrm[\string##1]}}}
\nolabels

\newdimen\fullhsize
\newdimen\hstitle
\hstitle=\hsize 
\newdimen\hsbody
\hsbody=\hsize 
\newdimen\hbodyoffset
\hbodyoffset=\hoffset 
\newbox\leftpage
\def\abstract#1{#1}
\def\rotated{\special{ps: landscape}
\magnification=1000  
\baselineskip=14pt
\global\hstitle=9truein\global\hsbody=4.75truein
\global\vsize=7truein\global\voffset=-.31truein
\global\hoffset=-0.54in\global\hbodyoffset=-.54truein
\global\fullhsize=10truein
\def\DefineTeXgraphics{%
\special{ps::[global]
/TeXgraphics {currentpoint translate 0.7 0.7 scale
              -80 0.72 mul -1000 0.72 mul translate} def}}
\let\lr=L
\def\ifsmall{\iftrue}
\def\titlepagefont{\twelvepoint}
\trueseventeenpoint
\def\almostshipout##1{\if L\lr \count1=1
      \global\setbox\leftpage=##1 \global\let\lr=R
   \else \count1=2
      \shipout\vbox{\hbox to\fullhsize{\box\leftpage\hfil##1}}
      \global\let\lr=L\fi}

\output={\ifnum\count0=1 
 \shipout\vbox{\hbox to \fullhsize{\hfill\pagebody\hfill}}\advancepageno
 \else
 \almostshipout{\leftline{\vbox{\pagebody\makefootline}}}\advancepageno
 \fi}

\def\abstract##1{{\leftskip=1.5in\rightskip=1.5in ##1\par}} }

\def\linemessage#1{\immediate\write16{#1}}

\global\newcount\secno \global\secno=0
\global\newcount\appno \global\appno=0
\global\newcount\meqno \global\meqno=1
\global\newcount\subsecno \global\subsecno=0
\global\newcount\figno \global\figno=0

\newif\ifAnyCounterChanged
\let\terminator=\relax
\def\normalize#1{\ifx#1\terminator\let\next=\relax\else%
\if#1i\aftergroup i\else\if#1v\aftergroup v\else\if#1x\aftergroup x%
\else\if#1l\aftergroup l\else\if#1c\aftergroup c\else%
\if#1m\aftergroup m\else%
\if#1I\aftergroup I\else\if#1V\aftergroup V\else\if#1X\aftergroup X%
\else\if#1L\aftergroup L\else\if#1C\aftergroup C\else%
\if#1M\aftergroup M\else\aftergroup#1\fi\fi\fi\fi\fi\fi\fi\fi\fi\fi\fi\fi%
\let\next=\normalize\fi%
\next}
\def\makeNormal#1#2{\def\doNormalDef{\edef#1}\begingroup%
\aftergroup\doNormalDef\aftergroup{\normalize#2\terminator\aftergroup}%
\endgroup}

\def\warnIfChanged#1#2{%
\ifundef#1
\else\begingroup%
\edef\oldDefinitionOfCounter{#1}\edef\newDefinitionOfCounter{#2}%
\ifx\oldDefinitionOfCounter\newDefinitionOfCounter%
\else%
\linemessage{Warning: definition of \noexpand#1 has changed.}%
\global\AnyCounterChangedtrue\fi\endgroup\fi}

\def\Section#1{\global\advance\secno by1\relax\global\meqno=1%
\global\subsecno=0%
\bigbreak\bigskip
\centerline{\twelvepoint \bf %
\the\secno. #1}%
\par\nobreak\medskip\nobreak}
\def\tagsection#1{%
\warnIfChanged#1{\the\secno}%
\xdef#1{\the\secno}%
\ifWritingAuxFile\immediate\write\auxfile{\noexpand\xdef\noexpand#1{#1}}\fi%
}
\def\section{\Section}
\def\Subsection#1{\global\advance\subsecno by1\relax\medskip %
\leftline{\bf\the\secno.\the\subsecno\ #1}%
\par\nobreak\smallskip\nobreak}
\def\tagsubsection#1{%
\warnIfChanged#1{\the\secno.\the\subsecno}%
\xdef#1{\the\secno.\the\subsecno}%
\ifWritingAuxFile\immediate\write\auxfile{\noexpand\xdef\noexpand#1{#1}}\fi%
}

\def\subsection{\Subsection}

\def\romappno{\uppercase\expandafter{\romannumeral\appno}}
\def\makeNormalizedRomappno{%
\expandafter\makeNormal\expandafter\normalizedromappno%
\expandafter{\romannumeral\appno}%
\edef\normalizedromappno{\uppercase{\normalizedromappno}}}
\def\Appendix#1{\global\advance\appno by1\relax\global\meqno=1\global\secno=0%
\global\subsecno=0%
\bigbreak\bigskip
\centerline{\twelvepoint \bf Appendix %
\romappno. #1}%
\par\nobreak\medskip\nobreak}
\def\tagappendix#1{\makeNormalizedRomappno%
\warnIfChanged#1{\normalizedromappno}%
\xdef#1{\normalizedromappno}%
\ifWritingAuxFile\immediate\write\auxfile{\noexpand\xdef\noexpand#1{#1}}\fi%
}
\def\appendix{\Appendix}
\def\Subappendix#1{\global\advance\subsecno by1\relax\medskip %
\leftline{\bf\romappno.\the\subsecno\ #1}%
\par\nobreak\smallskip\nobreak}
\def\tagsubappendix#1{\makeNormalizedRomappno%
\warnIfChanged#1{\normalizedromappno.\the\subsecno}%
\xdef#1{\normalizedromappno.\the\subsecno}%
\ifWritingAuxFile\immediate\write\auxfile{\noexpand\xdef\noexpand#1{#1}}\fi%
}

\def\eqn#1{\makeNormalizedRomappno%
\ifnum\secno>0%
  \warnIfChanged#1{\the\secno.\the\meqno}%
  \eqno(\the\secno.\the\meqno)\xdef#1{\the\secno.\the\meqno}%
     \global\advance\meqno by1
\else\ifnum\appno>0%
  \warnIfChanged#1{\normalizedromappno.\the\meqno}%
  \eqno({\rm\romappno}.\the\meqno)%
      \xdef#1{\normalizedromappno.\the\meqno}%
     \global\advance\meqno by1
\else%
  \warnIfChanged#1{\the\meqno}%
  \eqno(\the\meqno)\xdef#1{\the\meqno}%
     \global\advance\meqno by1
\fi\fi%
\eqlabeL#1%
\ifWritingAuxFile\immediate\write\auxfile{\noexpand\xdef\noexpand#1{#1}}\fi%
}
\def\defeqn#1{\makeNormalizedRomappno%
\ifnum\secno>0%
  \warnIfChanged#1{\the\secno.\the\meqno}%
  \xdef#1{\the\secno.\the\meqno}%
     \global\advance\meqno by1
\else\ifnum\appno>0%
  \warnIfChanged#1{\normalizedromappno.\the\meqno}%
  \xdef#1{\normalizedromappno.\the\meqno}%
     \global\advance\meqno by1
\else%
  \warnIfChanged#1{\the\meqno}%
  \xdef#1{\the\meqno}%
     \global\advance\meqno by1
\fi\fi%
\eqlabeL#1%
\ifWritingAuxFile\immediate\write\auxfile{\noexpand\xdef\noexpand#1{#1}}\fi%
}
\def\anoneqn{\makeNormalizedRomappno%
\ifnum\secno>0
  \eqno(\the\secno.\the\meqno)%
     \global\advance\meqno by1
\else\ifnum\appno>0
  \eqno({\rm\normalizedromappno}.\the\meqno)%
     \global\advance\meqno by1
\else
  \eqno(\the\meqno)%
     \global\advance\meqno by1
\fi\fi%
}
\def\mfig#1#2{\global\advance\figno by1%
\relax#1\the\figno%
\warnIfChanged#2{\the\figno}%
\edef#2{\the\figno}%
\reflabeL#2%
\ifWritingAuxFile\immediate\write\auxfile{\noexpand\xdef\noexpand#2{#2}}\fi%
}

\catcode`@=11 

\font\ninerm=cmr9
\font\eightrm=cmr8
\font\sixrm=cmr6

\def\loadtrueseventeenpoint{
 \font\seventeenrm=cmr10 at 17.28truept
 \font\seventeeni=cmmi10 at 17.28truept
 \font\seventeenbf=cmbx10 at 17.28truept
 \font\seventeenit=cmti10 at 17.28truept
 \font\seventeensl=cmsl10 at 17.28truept
 \font\seventeensy=cmsy10 at 17.28truept
}
\def\loadfourteenpoint{
\font\fourteenrm=cmr10 at 14.4pt
\font\fourteeni=cmmi10 at 14.4pt
\font\fourteenit=cmti10 at 14.4pt
\font\fourteensl=cmsl10 at 14.4pt
\font\fourteensy=cmsy10 at 14.4pt
\font\fourteenbf=cmbx10 at 14.4pt
}
\def\loadtruetwelvepoint{
\font\twelverm=cmr10 at 12truept
\font\twelvei=cmmi10 at 12truept
\font\twelveit=cmti10 at 12truept
\font\twelvesl=cmsl10 at 12truept
\font\twelvesy=cmsy10 at 12truept
\font\twelvebf=cmbx10 at 12truept
}

\font\ninei=cmmi9
\font\eighti=cmmi8
\font\sixi=cmmi6
\skewchar\ninei='177 \skewchar\eighti='177 \skewchar\sixi='177

\font\ninesy=cmsy9
\font\eightsy=cmsy8
\font\sixsy=cmsy6
\skewchar\ninesy='60 \skewchar\eightsy='60 \skewchar\sixsy='60

\font\ninebf=cmbx9
\font\eightbf=cmbx10
\font\sixbf=cmbx6

\font\ninett=cmtt9
\font\eighttt=cmtt8

\hyphenchar\tentt=-1 
\hyphenchar\ninett=-1
\hyphenchar\eighttt=-1

\font\ninesl=cmsl9
\font\eightsl=cmsl8

\font\nineit=cmti9
\font\eightit=cmti8


\newskip\ttglue
\def\tenpoint{\def\rm{\fam0\tenrm}%
  \textfont0=\tenrm \scriptfont0=\sevenrm \scriptscriptfont0=\fiverm
  \textfont1=\teni \scriptfont1=\seveni \scriptscriptfont1=\fivei
  \textfont2=\tensy \scriptfont2=\sevensy \scriptscriptfont2=\fivesy
  \textfont3=\tenex \scriptfont3=\tenex \scriptscriptfont3=\tenex
  \def\it{\fam\itfam\tenit}\textfont\itfam=\tenit
  \def\sl{\fam\slfam\tensl}\textfont\slfam=\tensl
  \def\bf{\fam\bffam\tenbf}\textfont\bffam=\tenbf \scriptfont\bffam=\sevenbf
  \scriptscriptfont\bffam=\fivebf
  \normalbaselineskip=12pt
  \let\sc=\eightrm
  \let\big=\tenbig
  \setbox\strutbox=\hbox{\vrule height8.5pt depth3.5pt width\z@}%
  \normalbaselines\rm}

\def\twelvepoint{\def\rm{\fam0\twelverm}%
  \textfont0=\twelverm \scriptfont0=\ninerm \scriptscriptfont0=\sevenrm
  \textfont1=\twelvei \scriptfont1=\ninei \scriptscriptfont1=\seveni
  \textfont2=\twelvesy \scriptfont2=\ninesy \scriptscriptfont2=\sevensy
  \textfont3=\tenex \scriptfont3=\tenex \scriptscriptfont3=\tenex
  \def\it{\fam\itfam\twelveit}\textfont\itfam=\twelveit
  \def\sl{\fam\slfam\twelvesl}\textfont\slfam=\twelvesl
  \def\bf{\fam\bffam\twelvebf}\textfont\bffam=\twelvebf%
  \scriptfont\bffam=\ninebf
  \scriptscriptfont\bffam=\sevenbf
  \normalbaselineskip=12pt
  \let\sc=\eightrm
  \let\big=\tenbig
  \setbox\strutbox=\hbox{\vrule height8.5pt depth3.5pt width\z@}%
  \normalbaselines\rm}

\def\fourteenpoint{\def\rm{\fam0\fourteenrm}%
  \textfont0=\fourteenrm \scriptfont0=\tenrm \scriptscriptfont0=\sevenrm
  \textfont1=\fourteeni \scriptfont1=\teni \scriptscriptfont1=\seveni
  \textfont2=\fourteensy \scriptfont2=\tensy \scriptscriptfont2=\sevensy
  \textfont3=\tenex \scriptfont3=\tenex \scriptscriptfont3=\tenex
  \def\it{\fam\itfam\fourteenit}\textfont\itfam=\fourteenit
  \def\sl{\fam\slfam\fourteensl}\textfont\slfam=\fourteensl
  \def\bf{\fam\bffam\fourteenbf}\textfont\bffam=\fourteenbf%
  \scriptfont\bffam=\tenbf
  \scriptscriptfont\bffam=\sevenbf
  \normalbaselineskip=17pt
  \let\sc=\elevenrm
  \let\big=\tenbig
  \setbox\strutbox=\hbox{\vrule height8.5pt depth3.5pt width\z@}%
  \normalbaselines\rm}

\def\seventeenpoint{\def\rm{\fam0\seventeenrm}%
  \textfont0=\seventeenrm \scriptfont0=\fourteenrm \scriptscriptfont0=\tenrm
  \textfont1=\seventeeni \scriptfont1=\fourteeni \scriptscriptfont1=\teni
  \textfont2=\seventeensy \scriptfont2=\fourteensy \scriptscriptfont2=\tensy
  \textfont3=\tenex \scriptfont3=\tenex \scriptscriptfont3=\tenex
  \def\it{\fam\itfam\seventeenit}\textfont\itfam=\seventeenit
  \def\sl{\fam\slfam\seventeensl}\textfont\slfam=\seventeensl
  \def\bf{\fam\bffam\seventeenbf}\textfont\bffam=\seventeenbf%
  \scriptfont\bffam=\fourteenbf
  \scriptscriptfont\bffam=\twelvebf
  \normalbaselineskip=21pt
  \let\sc=\fourteenrm
  \let\big=\tenbig
  \setbox\strutbox=\hbox{\vrule height 12pt depth 6pt width\z@}%
  \normalbaselines\rm}

\def\ninepoint{\def\rm{\fam0\ninerm}%
  \textfont0=\ninerm \scriptfont0=\sixrm \scriptscriptfont0=\fiverm
  \textfont1=\ninei \scriptfont1=\sixi \scriptscriptfont1=\fivei
  \textfont2=\ninesy \scriptfont2=\sixsy \scriptscriptfont2=\fivesy
  \textfont3=\tenex \scriptfont3=\tenex \scriptscriptfont3=\tenex
  \def\it{\fam\itfam\nineit}\textfont\itfam=\nineit
  \def\sl{\fam\slfam\ninesl}\textfont\slfam=\ninesl
  \def\bf{\fam\bffam\ninebf}\textfont\bffam=\ninebf \scriptfont\bffam=\sixbf
  \scriptscriptfont\bffam=\fivebf
  \normalbaselineskip=11pt
  \let\sc=\sevenrm
  \let\big=\ninebig
  \setbox\strutbox=\hbox{\vrule height8pt depth3pt width\z@}%
  \normalbaselines\rm}

\def\eightpoint{\def\rm{\fam0\eightrm}%
  \textfont0=\eightrm \scriptfont0=\sixrm \scriptscriptfont0=\fiverm%
  \textfont1=\eighti \scriptfont1=\sixi \scriptscriptfont1=\fivei%
  \textfont2=\eightsy \scriptfont2=\sixsy \scriptscriptfont2=\fivesy%
  \textfont3=\tenex \scriptfont3=\tenex \scriptscriptfont3=\tenex%
  \def\it{\fam\itfam\eightit}\textfont\itfam=\eightit%
  \def\sl{\fam\slfam\eightsl}\textfont\slfam=\eightsl%
  \def\bf{\fam\bffam\eightbf}\textfont\bffam=\eightbf \scriptfont\bffam=\sixbf%
  \scriptscriptfont\bffam=\fivebf%
  \normalbaselineskip=9pt%
  \let\sc=\sixrm%
  \let\big=\eightbig%
  \setbox\strutbox=\hbox{\vrule height7pt depth2pt width\z@}%
  \normalbaselines\rm}

\def\tenbig#1{{\hbox{$\left#1\vbox to8.5pt{}\right.\n@space$}}}
\def\ninebig#1{{\hbox{$\textfont0=\tenrm\textfont2=\tensy
  \left#1\vbox to7.25pt{}\right.\n@space$}}}
\def\eightbig#1{{\hbox{$\textfont0=\ninerm\textfont2=\ninesy
  \left#1\vbox to6.5pt{}\right.\n@space$}}}

\def\footnote#1{\edef\@sf{\spacefactor\the\spacefactor}#1\@sf
      \insert\footins\bgroup\eightpoint
      \interlinepenalty100 \let\par=\endgraf
        \leftskip=\z@skip \rightskip=\z@skip
        \splittopskip=10pt plus 1pt minus 1pt \floatingpenalty=20000
        \smallskip\item{#1}\bgroup\strut\aftergroup\@foot\let\next}
\skip\footins=12pt plus 2pt minus 4pt 
\dimen\footins=30pc 

\newinsert\margin
\dimen\margin=\maxdimen
\def\titlefont{\seventeenpoint}
\loadtruetwelvepoint 
\loadtrueseventeenpoint

\def\eatOne#1{}
\def\ifundef#1{\expandafter\ifx%
\csname\expandafter\eatOne\string#1\endcsname\relax}
\def\notTrue{\iffalse}\def\isTrue{\iftrue}
\def\ifdef#1{{\ifundef#1%
\aftergroup\notTrue\else\aftergroup\isTrue\fi}}
\def\use#1{\ifundef#1\linemessage{Warning: \string#1 is undefined.}%
{\tt \string#1}\else#1\fi}


\global\newcount\refno \global\refno=1
\newwrite\rfile
\newlinechar=`\^^J
\def\@ref#1#2{\the\refno\n@ref#1{#2}}
\def\n@ref#1#2{\xdef#1{\the\refno}%
\ifnum\refno=1\immediate\openout\rfile=\jobname.refs\fi%
\immediate\write\rfile{\noexpand\item{[\noexpand#1]\ }#2.}%
\global\advance\refno by1}
\def\nref{\n@ref} 
\def\ref{\@ref}   
\def\lref#1#2{\the\refno\xdef#1{\the\refno}%
\ifnum\refno=1\immediate\openout\rfile=\jobname.refs\fi%
\immediate\write\rfile{\noexpand\item{[\noexpand#1]\ }#2\semi}%
\global\advance\refno by1}
\def\cref#1{\immediate\write\rfile{#1\semi}}

\def\preref#1#2{\gdef#1{\@ref#1{#2}}}

\def\semi{;\hfil\noexpand\break}

\def\listrefs{\vfill\eject\immediate\closeout\rfile
\centerline{{\bf References}}\bigskip\frenchspacing%
\input \jobname.refs\vfill\eject\nonfrenchspacing}

\def\inputAuxIfPresent#1{\immediate\openin1=#1
\ifeof1\message{No file \auxfileName; I'll create one.
}\else\closein1\relax\input\auxfileName\fi%
}
\def\NPB{Nucl.\ Phys.\ B}

\newif\ifWritingAuxFile
\newwrite\auxfile
\def\SetUpAuxFile{%
\xdef\auxfileName{\jobname.aux}%
\inputAuxIfPresent{\auxfileName}%
\WritingAuxFiletrue%
\immediate\openout\auxfile=\auxfileName}


\def\RP{\right.}


\catcode`\@=\active
\catcode`@=12  
\catcode`\"=\active

\SetUpAuxFile
\loadfourteenpoint

\nopagenumbers\hsize=\hstitle\vskip1in
\overfullrule 0pt
\hfuzz 35 pt
\vbadness=10001

\def\"#1{{\accent127 #1}}

\def\coeff#1#2{{\textstyle{#1\over#2}}}
\def\frac#1#2{{#1\over#2}}
\def\hf{\coeff12}
\def\lr{\leftrightarrow}

\def\Ord{{\cal O}}

\def\Atree{A^{\rm tree}}
\def\Aloop{A^{\rm 1-loop}}
\def\Re{{\rm Re}\,}
\def\Im{{\rm Im}\,}
\def\as{\alpha_s}
\def\sstw{\sin^2\theta_W}
\def\cstw{\cos^2\theta_W}
\def\ctw{\cos\theta_W}
\def\qb{\bar{q}}
\def\xb{\bar{x}}
\def\thnormal{\theta_n^{(q\qb)}}
\def\costh{\cos\theta_n}
\def\costetn{\langle\cos\theta_n\rangle}
\def\Pe{P_e}
\def\Li{\mathop{\rm Li}\nolimits}
\def\ycut{y_{\rm cut}}
\def\qcd{{\rm QCD}}
\def\hky{Zg^*g}
\def\ew{{\rm EW}}

\def\xps{x_+^s}
\def\xms{x_-^s}
\def\xpt{x_+^t}
\def\xmt{x_-^t}
\def\lrho{\ell^{\rho}}
%

\noindent
hep-ph/9505355 \hfill {SLAC--PUB--6725}\break
\rightline{PITHA 94/62}
\rightline{May, 1995}
\rightline{(M)}
\rightline{   }

\leftlabelstrue
\vskip -1.0 in
\Title{Event Handedness in $e^+e^-$ Annihilation to
Three Jets${}^{\star}$}

\centerline{Arnd Brandenburg${}^{\dagger}$}
\vskip 0.05truein
\baselineskip12truept
\centerline{\it Institut f\"ur Theoretische Physik}
\centerline{\it Physikzentrum}
\centerline{\it Rheinisch-Westf\"alische Technische Hochschule Aachen}
\centerline{\it 52056 Aachen, Germany}

\smallskip\smallskip

\baselineskip17truept
\centerline{Lance Dixon and Yael Shadmi${}^{\ddagger}$}
\vskip 0.05truein
\baselineskip12truept
\centerline{\it Stanford Linear Accelerator Center}
\centerline{\it Stanford University}
\centerline{\it Stanford, CA 94309}
\vskip 0.2in\baselineskip13truept

\vskip 0.5truein
\centerline{\bf Abstract}

{\narrower

We discuss rescattering effects that can be measured
in $e^+e^-$ annihilation to three jets through a single gauge boson,
by using triple product (``event handedness'')
correlations of the $Z$ ($\gamma^*$)
polarization with jet momenta.
The gauge boson polarization may be produced either by polarized beams
or through the natural polarization (left-right asymmetry) of the $Z$.
QCD rescattering does not
generate  triple product correlations at one loop for
massless quarks. We therefore calculate the QCD contribution
for massive quarks, as well as the
contribution of $W$ and $Z$ exchange loops for massless quarks.
Due to various cancellations, the standard model predictions for
triple-product correlations at the $Z$ are very small,
making such measurements potentially sensitive
to physics beyond the standard model.
For example, the effects of a new gauge boson that couples only to baryon
number may be larger than the standard model contributions; however
the effects would probably still be too small to effectively constrain it.}

\vskip 0.3truein

\centerline{\sl Submitted to Physical Review D}

\vfill
\vskip 0.1in
\noindent\hrule width 3.6in\hfil\break
\noindent
${}^{\star}$Research supported by the US Department of
Energy under grant DE-AC03-76SF00515.\hfil\break
${}^{\dagger}$Research supported in part by the Max Kade
Foundation.\hfil\break
${}^{\ddagger}$Present address: Fermi National Accelerator Laboratory,
P.O. Box 500, Batavia, IL 60510.\hfil\break

\Date{}

\line{}

\baselineskip17pt
%


\preref\Nachtmann{%
W. Bernreuther, U. L\"{o}w, J.P. Ma and O. Nachtmann, Z.\ Phys.\
C43:117 (1989)}

\preref\cpcollect{%
For example:
J. F. Donoghue and G. Valencia, Phys.\ Rev.\ Lett. 58:451 (1987);
M. B. Gavela, F. Iddir, A. Le Yaouanc, L. Oliver, O. Pene and J. C.
Raynal, Phys. Rev. D39:1870 (1989);
J. Bernab\'eu and N. Rius, Phys. Lett. B232:127 (1989);
M. Kamionkowski, Phys. Rev. D41:1672 (1990);
D. Atwood, S. Bar-Shalom and A. Soni, Phys.\ Rev.\ D51:1034 (1995)}

\preref\BILAL{%
A.Bilal, E. Mass\'o and A. De R\'ujula, \NPB 355:549 (1991)
}

\preref\FKSS{%
K. Fabricius, G. Kramer, G. Schierholz and I. Schmitt,
Phys.\ Rev.\ Lett. 45:867 (1980)}

\preref\SLDALR{%
K. Abe et al., Phys.\ Rev.\ Lett. 73:25 (1994)}

\preref\LEP{%
ALEPH Collaboration, DELPHI Collaboration,
L3 Collaboration, OPAL Collaboration; LEP Electroweak Working Group,
CERN-PPE-94-187}

\preref\jethand{%
O. Nachtmann, Nucl.\ Phys.\ B127:314 (1977)\semi
A.V. Efremov, Sov.\ J.\ Nucl.\ Phys.\ 28:83 (1978)\semi
R.H. Dalitz, G.R. Goldstein and R. Marshall, Z.\ Phys.\ C42:441
(1989)\semi
A.V. Efremov, L. Mankiewicz and N.A. T\"{o}rnqvist, Phys.\ Lett.\
B284:394 (1992)}

\preref\DKD{%
A. De R\'ujula, J.M. Kaplan and E. de Rafael,
Nucl.\ Phys.\ B35:365 (1971)\semi
A. De R\'ujula, R. Petronzio and B. Lautrup, Nucl.\ Phys.\ B146:50
(1978)}

\preref\RP{%
J.P. Ralston and B. Pire, in {\it Proc. of 5th Intl. Symp. on
High Energy Spin Physics, Upton, N.Y.}, published in
BNL Spin Sympos. (1982);
Phys.\ Rev.\ D28:260 (1983)}

\preref\CW{%
R. Carlitz and R. Willey, Phys.\ Rev.\ D45:2323 (1992)}

\preref\HHKe{%
K. Hagiwara, K. Hikasa and N. Kai, Phys.\ Rev.\ D27:84 (1983)}

\preref\HHKnu{%
K. Hagiwara, K. Hikasa and N. Kai, Phys.\ Rev.\ Lett. 47:983 (1981)}

\preref\FKKSS{%
K. Fabricius, J.G. K\"{o}rner, G. Kramer, G. Schierholz and I. Schmitt,
Phys.\ Lett.\ 94B:207 (1980)}

\preref\KoernerSchuler{%
J.G. K\"{o}rner and G. Schuler, Z.\ Phys.\ C26:559 (1985)}

\preref\KUNSZT{%
Z. Kunszt, A. Signer and Z. Tr\'ocs\'anyi, \NPB 420:550 (1994)}

\preref\HKY{%
K. Hagiwara, T. Kuruma and Y. Yamada, Nucl.\ Phys.\ B358:80 (1991)}

\preref\CM{%
C. Carone and H. Murayama, Phys.\ Rev.\ Lett. 74:3122 (1995);
Phys.\ Rev.\ D52:484 (1995), hep-ph/9501220}

\preref\BD{%
D. Bailey and S. Davidson, Phys.\ Lett.\ 348B:185 (1995)}

\preref\bratkov{%
E.L. Bratkovskaya, E.A. Kuraev, Z.K. Silagadze and O.V. Terayaev,
Phys.\ Lett.\ B338:471 (1994), hep-ph/9412230}

\preref\OurIntegrals{%
Z. Bern, L. Dixon and D.A. Kosower, Phys.\ Lett.\ 302B:299 (1993);
erratum {\it ibid.} 318:649 (1993); \NPB 412:751 (1994)}

\preref\OPT{%
D. Atwood and A. Soni, Phys.\ Rev.\ D45:2405 (1992)}

\preref\stav{%
H. Olsen and J. Stav, Phys.\ Rev.\ D50:6775 (1994)}

\preref\SLDbound{%
K. Abe et al., preprint SLAC--PUB--6969, to be published in Phys.\ Rev.\
Lett}


$\null$
\vskip -3. cm

\section{Introduction}
\tagsection\IntroSection

The standard model has withstood experimental scrutiny
remarkably well, even as precision measurements at LEP, SLC and
the Tevatron are becoming sensitive to electroweak radiative
corrections.
It is important to test the standard model with as many observables
as possible.  Observables that vanish identically at tree level
are special, in that {\it any} nonzero measurement of such a
quantity simultaneously probes higher-order standard model
corrections and potential physics beyond the standard model.
Examples of such observables include the GIM-protected processes
$K^0 \to \bar{K}^0$ and $b\to s\gamma$, as well as many CP-violating
quantities.

It is also possible to construct tree-vanishing observables in jet
physics.
Consider the following observable in $e^+e^-$ annihilation
into three jets,
$$
 {\bf \hat{k}_e \cdot (k_1 \times k_2) } ,
\eqn\tripleprod
$$
where ${\bf k_1}$ and ${\bf k_2}$ are the momentum
vectors of jets 1 and 2, labeled according to the energy-ordering
$E_1 > E_2 > E_3$, and ${\bf \hat{k}_e}$ is the electron beam direction.
A  triple product correlation may be defined as the expectation value
of~(\use\tripleprod),
$$
\langle {\bf \hat{k}_e \cdot (k_1 \times k_2) } \rangle.
\eqn\triplecorr
$$
On the $Z$ pole,~(\use\triplecorr) is proportional
to the $Z$ boson polarization, which
may be produced either with
longitudinally polarized beams
(such as the $\sim 60-80\%$ polarized electrons available at
SLC~[\use\SLDALR]), or with unpolarized
beams~[\use\LEP],
utilizing the natural $Z$ polarization induced by the left-right
asymmetry $A_{LR}^{(e)} \sim 14\%$~[\use\SLDALR].
In both cases the polarization vector of the $Z$ boson sample
points along the beam direction.

In the case of $e^+e^-$
annihilation into a virtual photon, i.e. in the absence
of axial vector couplings, one needs
longitudinally polarized beams to get a nonzero
value of~(\use\triplecorr). This case was first discussed
in~[\use\FKKSS,\use\FKSS].

The observable~(\use\tripleprod) is even under CP, and odd under
T$_{\rm N}$, where T$_{\rm N}$ reverses spatial momenta and spin
vectors.  T$_{\rm N}$ does not exchange initial and final states, and
so it is not the true time reversal operation T.
Because of the distinction between T$_{\rm N}$ and T, a nonzero
value of~(\use\triplecorr) does {\it not} signal CPT violation.
It can be produced by final-state rescattering, even in a theory
that respects CP and T~[\use\DKD].

There are other variations on~(\use\tripleprod),
with the same symmetry properties, which will be discussed in due course.
In particular, instead of the expectation value~(\use\triplecorr), one
can discuss the asymmetry associated with it, namely
$$
{
{\rm N}\,({\bf \hat{k}_e \cdot (k_1 \times k_2) }>0)\ -\
{\rm N}\,({\bf \hat{k}_e \cdot (k_1 \times k_2) }<0)\ \over
{\rm N}\,({\bf \hat{k}_e \cdot (k_1 \times k_2) }>0)\ +\
{\rm N}\,({\bf \hat{k}_e \cdot (k_1 \times k_2) }<0)\ }\ .
\eqn\asymm
$$
where ${\rm N}\,({\bf \hat{k}_e \cdot (k_1 \times k_2) }>0)$ is the number
of three jet events for which ${\bf \hat{k}_e \cdot (k_1 \times k_2) }>0$,
etc. In the following, we will use the terms triple product correlation and
asymmetry interchangeably.

The large number of polarized $Z$ bosons now available, at both SLC and LEP,
allows for sensitive tests of rescattering effects, through measurement of
triple product correlations such as~(\use\triplecorr).
It is therefore important to calculate the standard model predictions,
and that is the main goal of this work.

Triple product correlations in $e^+e^-$ annihilation into jets were
also proposed for the study of CP violation~[\use\cpcollect,\use\Nachtmann]
and (in $e^+e^-\to W^+W^-$) for the study of weak gauge boson
couplings~[\use\BILAL].

The triple product correlation~(\use\triplecorr) could also
be termed ``event handedness'', by analogy to ``jet handedness''
observables~[\use\jethand] in which ${\bf \hat{k}_e}$ is replaced by
the axis of a jet produced by a longitudinally polarized parton,
and ${\bf k_i}$ become momenta of particles inside
that single jet, rather than jet momenta.
At the event level, as opposed to the jet level, one probes
rescattering phases generated at much shorter distance scales,
where perturbative techniques may be applied.

In a covariant framework, a nonzero triple product
correlation in $e^+e^-\to$ three jets is produced by terms in the
differential cross-section that are proportional to
the Levi-Civita tensor $\varepsilon_{\mu\nu\sigma\rho}$
contracted with four of the five momentum vectors in the problem.
(Up to a sign, different choices of the four momenta give
the same contraction, due to momentum conservation.)
The contracted Levi-Civita tensor must be multiplied by the imaginary
part of some loop integral, in order to contribute to the
differential cross-section.

A first guess for how a triple product correlation might be
generated in the standard model is via QCD rescattering of the
final-state partons in $e^+e^-\to (\gamma^*,Z) \to qg\qb$.
Indeed, in crossed channels such as
$q\qb \to g(\gamma^*,Z) \to g\ell^+\ell^-$,
which contribute to Drell-Yan production in (polarized) proton-proton
scattering, it has been shown~[\use\RP,\use\CW] that one loop QCD
generates a nonvanishing single spin asymmetry,
very much like~(\triplecorr).
Similar effects occur in semi-inclusive deep-inelastic scattering,
$e^-p \to e^-hX$, where $h$ is a single hadron~[\use\HHKe].

Amusingly, however, rescattering effects in QCD with massless quarks
do {\it not} generate the triple product correlation at one loop.
The various cuts conspire to precisely
cancel each other in the fully time-like kinematics of $e^+e^-$
annihilation through a vector boson, unless some
of the particles propagating around the loop are
massive~[\use\HHKe,\use\FKKSS].
As we show in
appendix~I,
generalizing an argument due to K\"{o}rner and
Schuler~[\use\KoernerSchuler],
this vanishing holds for
$e^+e^-\to (\gamma^*,Z) \to n$-partons at one loop.
Though we do not have a proof, we expect --- and we will
assume here --- that the argument goes through to all orders in
$\alpha_s$ perturbation theory for massless partons.
We expect the argument to break down at the nonperturbative level,
due to the dynamical generation of particle masses in QCD at this
level.

In the standard model, there are three possible sources for the particle
masses needed to generate the 3-jet triple product correlation:

1) In QCD rescattering, one can include the effects of nonzero
quark masses, in diagrams such as figure~1.
These effects were first calculated by Fabricius,
Kramer, Schierholz and Schmitt~[\use\FKSS] in the case of a virtual photon
(no quark axial coupling contributions); they presented numerical
results for two choices of
$m_q/\sqrt{s}$.
At the $Z$ peak, the only significant quark mass is $m_b$.
A naive estimate of the size of the triple product~(\triplecorr)
generated by the $b$ quark mass in $e^+e^- \to b\bar{b}g$ is
$$
  N_c \alpha_s {m_b^2 \over M_Z^2} M_Z^2 \approx 10^{-3} M_Z^2,
\eqn\qcdestimate
$$
where $N_c=3$ is a color factor, $\alpha_s$ comes from the additional
strong coupling constants, beyond those present in the tree-level 3-jet
production rate, and ${m_b^2 \over M_Z^2}$ reflects the fact that
the effect must vanish as $m_b \to 0$.  (It cannot
vanish as $m_b$, since the suppression is a kinematic effect,
independent of whether the $b$ is a fermion or a scalar.)
Of course the corresponding contribution would be even smaller for
$u,d,s,c$.

2) There is another type of QCD ``rescattering'' where the massive
quark annihilates and is not an external state, first studied by
Hagiwara, Kuruma and Yamada~[\use\HKY].  This contribution
requires a triangle diagram with two external
gluons (see figure~2); due to Furry's theorem the third vector boson
must have an axial coupling to the quark in the triangle loop,
i.e. it must be the $Z$ rather than the photon.
Naively this contribution is of the
same order~(\qcdestimate), except that it lacks the factor of $N_c$.
Also, it can contribute to non-$b$ final states, so one might expect
a compensating factor of
$n_f=5$.
However, the relative contributions of up- and down-type quarks in
the final state turn out to be opposite in sign and almost equal in
magnitude, so one does not get the $n_f$ enhancement, and we will see that
this contribution is much smaller than the first one at all energies below
the $t\bar{t}$ threshold.

3) A final possibility is electroweak rescattering, the
exchange of a $W$ or a $Z$ between the outgoing quark-antiquark pair
(see figure~3).  In this case the naive estimate is just
$$
  \alpha_W  M_Z^2 \approx {1\over30} M_Z^2\ .
\eqn\ewestimate
$$
Electroweak rescattering can only compete with QCD rescattering
because of the quark mass suppression in equation~(\qcdestimate).

As we show, all of these naive estimates turn out to be overestimates,
due largely to phase space factors, and the standard model
contributions are quite small.
We will also consider one beyond-the-standard-model effect that might
be seen or constrained by this observable.
In order to generate a triple-product asymmetry, at least two of
the particles propagating around the loop must be on-shell.
But if the new particles one wants to probe must be produced on-shell
to generate an asymmetry, it may be easier to constrain them based on
their direct production rather than by using the asymmetry. Thus, one does not
expect large contributions from supersymmetry effects. For example, one
loop diagrams involving squarks and gluinos would not contribute to a
three-jet asymmetry at the $Z$ pole, since the squark propagators would
be off-shell, given current bounds on squark masses.
The asymmetries may be more sensitive to the exchange of a single
new particle.
If a gauge boson $B$ couples to baryon number,
and therefore does not couple directly to leptons, then it is
hard to detect by other means even if it is as light as 10-20
GeV~[\use\CM,\use\BD].
Yet in this mass range it would give a result like the electroweak
result, except potentially scaled up by a large factor, if the
coupling constant $\alpha_B$ is larger than the electroweak coupling
constant and if the $B$ mass is significantly less than
the $W$ and $Z$ masses.
The contribution of this hypothetical $B$ boson is simply obtained from
the electroweak calculation.

Three types of standard model contributions to the triple
product~(\use\triplecorr) in $e^+e^-$ annihilation
are {\it not} investigated in this paper.
In the first two, the $e^+e^-$ annihilation does not proceed through a
single gauge boson, so the kinematic invariants appearing in the loop
integrals are not all timelike.  Therefore the argument in appendix~I
does not apply, and non-vanishing triple product correlations can be
generated even when all particle masses are set to zero.

\par\noindent
1) The electron-positron annihilation can produce a
$\gamma\gamma$ pair, or a $\gamma Z$ pair, which then rescatter into
the $q\qb g$ final state.  (See figure~4.)
This contribution is likely to be very small at the $Z$ pole, because
it is proportional to $\alpha_{QED}$ and one does not get the advantage
of the $Z$ pole (unless the photon in the $\gamma Z$ intermediate state
is very soft).
\par\noindent
2) Another possibility is two-photon physics, $\gamma\gamma\to q\qb g$,
where the photons are produced as initial-state radiation.
(See figure~5.)   At a real $\gamma\gamma$
collider the analogous triple product may be sizable.
In $e^+e^-$ annihilation at the $Z$ pole, however, the initial-state
radiation is likely to be too small to make this contribution
observable.
\par\noindent
3) All of the above contributions are those of short-distance physics.
In addition there may be long-distance, nonperturbative QCD
effects. (Such contributions to spin-momentum correlations
in $e^+e^-\to 4\pi$ were discussed in~[\use\bratkov].)
These should be suppressed by some power of $\Lambda_{QCD}/\sqrt{s}$,
but in the absence of an operator product expansion we do not know
the precise power, let alone the prefactor.  Some nonperturbative
effects can be estimated using a hadronization Monte Carlo, but this one
seems particularly difficult because of the need to keep track of
phases to get the effect.

Finally, we note that the $Z$ width, or more generally, imaginary parts
of vacuum polarization and vertex corrections in the leptonic part of
the cross-section, do not contribute to~(\use\triplecorr).
These only renormalize the
tree amplitude, and therefore, as in the case of the soft singularities
discussed in
appendix~I, cannot generate the triple product asymmetry.


\section{Notation}
\tagsection\NotationSection

The $e^+e^- \to (\gamma^*,Z) \to q\qb g$ differential cross-section
at center-of-mass energy $\sqrt{s}$, assuming no transverse beam
polarization, can be written as follows (we adopt the notation of
ref.~[\use\HKY]):
$$
\eqalign{
  {d^4\sigma \over dx\, d\xb\, d\cos\theta\, d\phi}\ =\
  {3\over4\pi}{\alpha_s\over\pi}\sigma_{\rm pt} \times \Bigl[
&   F_1 (1+\cos^2\theta) + F_2 (1-3\cos^2\theta) + F_3 \cos\theta \cr
& + F_4 \sin2\theta\cos\phi + F_5 \sin^2\theta \cos2\phi
  + F_6\sin\theta\cos\phi \cr
& + F_7 \sin2\theta\sin\phi + F_8 \sin^2\theta \sin2\phi
  + F_9\sin\theta\sin\phi \Bigr]\ , \cr }
\eqn\hagidecomp
$$
where
$$
 \sigma_{\rm pt}\ =\ \sigma(e^+e^- \to \gamma^* \to \mu^+\mu^-)
  \ =\ {4\pi\alpha^2\over 3s}\ .
\eqn\ptcrosssection
$$
The only kinematic variables appearing in the functions $F_i$
are the scaled quark and antiquark energies
in the $e^+e^-$ center-of-mass frame,
$x=2E_q/\sqrt{s}$ and $\xb=2E_{\qb}/\sqrt{s}$.
The angle between the electron direction and the quark direction
is $\theta$ (see figure~6), and the (signed) angle between the $e^+e^-q$ plane
and the $q\qb g$ plane is $\phi$.
Denote by $\thnormal$ the angle between the electron direction
and the normal to the $q\qb g$ event plane,
$$
  \cos\thnormal\ =\
  { {\bf \hat{k}_e} \cdot ({\bf k_q}\times {\bf k_{\bar{q}}})
    \over |{\bf k_q}\times {\bf k_{\bar{q}}}| }\ .
\eqn\thnormaldef
$$
This angle is related to $(\theta,\phi)$ by
$$
  \cos\thnormal\ =\ \sin\theta\sin\phi,
\eqn\thnormalrel
$$
so the observable~(\tripleprod) derives from the function $F_9$.
Functions $F_7$ and $F_8$ are also odd under T$_{\rm N}$,
but in a CP invariant theory they give vanishing contribution to
observables in which the quark and antiquark are not distinguished from
each other~[\use\Nachtmann], and so we will not consider them
further at this time.

The distribution in the normal angle $\thnormal$, after integrating
over the remaining angle, is
$$
\eqalign{
  {d^3\sigma \over dx\, d\xb\, d\cos\thnormal}\ =\
  {3\over2}{\alpha_s\over\pi}\sigma_{\rm pt} \times \Bigl[
&       F_1 (1+\hf\sin^2\thnormal)
    + (F_2-F_5) (1-\coeff{3}{2}\sin^2\thnormal) \cr
&   + F_9 \cos\thnormal \Bigr]\ , \cr }
\eqn\hagithnormal
$$

We assume that the electrons have longitudinal polarization $\Pe$,
with $\Pe = +1$ for right-handed electrons, while the positrons are
taken to be unpolarized, although polarized positrons can be treated
easily as well.

The denominator of the expectation value~(\triplecorr) is
found (to lowest order in $\alpha_s$) by integrating the tree
approximations to $F_1$ and $F_2-F_5$ over the Dalitz plot with some
three-jet cut, e.g. on the thrust $T$ or on the invariant masses of
parton pairs.

The tree-level approximations to $F_i$ are given by
$$
  F_i(x,\xb)\ =\ g_i^{(0),v} F_i^{(0),v}(x,\xb)
            \ +\ g_i^{(0),a} F_i^{(0),a}(x,\xb)\ +\ \Ord(\as),
\eqn\fiapprox
$$
where the tree-level coupling factors for P-even terms are
$$
\eqalign{
  g_i^{(0),v}\ &=\
  \Bigl( (V_e^2+A_e^2) -2V_eA_e \Pe \Bigr) V_q^2 |\chi(s)|^2
    \ -\ 2Q(V_e-A_e \Pe)V_q \Re \chi(s)\ +\ Q^2, \cr
  g_i^{(0),a}\ &=\
  \Bigl( (V_e^2+A_e^2) -2V_eA_e \Pe \Bigr) A_q^2 |\chi(s)|^2, \cr
  g_i^{(0)}\ &=\ g_i^{(0),v}+g_i^{(0),a} \cr
  \ &=\
  \Bigl( (V_e^2+A_e^2) -2V_eA_e \Pe \Bigr) (V_q^2+A_q^2)|\chi(s)|^2
    \ -\ 2Q(V_e-A_e \Pe)V_q \Re \chi(s)\ +\ Q^2, \cr
   &\qquad\qquad\qquad i=1,2,4,5.
\cr}
\eqn\treecouple
$$
Here $Q$ is the charge of the quark,
$V_e$, $A_e$, $V_q$, $A_q$ are the vector and axial-vector
couplings of the $Z$ to the electron and (external) quark,
$$\eqalign{
  V\ &=\ \hf I_3 - q \sstw \, , \qquad A\ =\ \hf I_3 \, , \cr
  \chi(s)\ &=\ {1\over\sstw\cstw}{s\over s-M_Z^2 + iM_Z\Gamma_Z} \ ,
\cr}
\eqn\vachidef
$$
with $I_3$ and $q$ the third component of isospin and charge
of the electron or quark.

At or around the $Z$ resonance, it is an excellent approximation
to set the external quark masses to zero in the tree cross-section. The
kinematic functions are then
$$
\eqalign{
  F_1^{(0),v}\ &=\ F_1^{(0),a}\ =\ { x^2+\xb^2 \over 2(1-x)(1-\xb) },
  \cr
  F_2^{(0),v}-F_5^{(0),v}\ &=\ F_2^{(0),a}-F_5^{(0),a}\ =\ 0, \cr
}\eqn\masslesstreef
$$
and only the coupling $g_1^{(0)}$ contributes in the denominator
of~(\use\triplecorr).
The full expressions, keeping quark masses,
which are needed to estimate the effects at
energies below the $Z$ resonance, are more complicated.
They are given in
appendix~II
and agree
with the results of~[\use\stav].


\section{Contributions to ${\bf F_9}$}
\tagsection\FnineSection

Now we compile the one loop contributions to $F_9(x,\xb)$ from the
various sources. All the contributions are proportional to
$$
|{\bf k_q}\times{\bf k_{\bar{q}}}|\ =\ {s\over 8}\,
\sqrt{(1-x)(1-\xb)(x+\xb-1)-z(2-x-\xb)^2}\ ,
\eqn\squareexpl
$$
so it is convenient to factor out the square-root appearing
in~(\squareexpl) from the expressions to follow.
Note that in the center-of-mass system, it does not matter
which two of the outgoing parton momenta appear on the left hand side
of~(\squareexpl).

\vskip0.3truein
\par\noindent
{\bf 1. QCD
${\bf e^+e^- \to (\gamma^*,Z) \to b\bar{b}g}$ Contribution \hfill }
\par\noindent
Write
$$
  F_{9,\qcd}(x,\xb)\ =\ {\alpha_s\over\pi}
  \sqrt{(1-x)(1-\xb)(x+\xb-1)-z(2-x-\xb)^2}
  \Bigl[ g_9^{(1),v} f^v_{\qcd} + g_9^{(1),a} f^a_{\qcd} \Bigr]\ ,
\eqn\fkssfnine
$$
where
$$
\eqalign{
  g_9^{(1),v}\ &=\ \Bigl( -2V_eA_e+(V_e^2+A_e^2)\Pe \Bigr)
   V_q^2 |\chi(s)|^2
  \ -\ 2Q(V_e\Pe - A_e)V_q \Re\chi(s)  \ +\ Q^2 \Pe, \cr
  g_9^{(1),a}\ &=\ \Bigl( -2V_eA_e+(V_e^2+A_e^2)\Pe \Bigr)
   A_q^2 |\chi(s)|^2. \cr
}\eqn\gnineone
$$
The functions $f^v_{\qcd}$, $f^a_{\qcd}$ can be decomposed into
leading-color and subleading-color contributions,
$$
\eqalign{
  f^v_{\qcd}\ &=\ 3\, f^{v,1} + \coeff{1}{3}\, f^{v,2}, \cr
  f^a_{\qcd}\ &=\ 3\, f^{a,1} + \coeff{1}{3}\, f^{a,2}, \cr
}\eqn\ncdecomp
$$
whose rather lengthy expressions are given in
formulae~(\use\coeffdec) through~(\use\rescoeffatwotwo)
of
appendix~II.

Numerical results for the
vector part of $F_{9,\qcd}$ (in the notation used here), as a function of the
thrust and the angle between quark and antiquark momenta,
were presented in ref.~[\use\FKSS] for two values of $m_q/\sqrt{s}$.
Our results have the opposite sign.
The absolute values presented in figure~3 of ref.~[\use\FKSS] also
differ slightly from ours.

\vskip0.3truein
\par\noindent
{\bf 2. QCD
${\bf e^+e^- \to Z \to g^*g \to q\qb g}$ Contribution \hfill }
\par\noindent
In this case the quarks in the loop may differ from the external quark;
denote the internal quark masses by $m_i$, where $i$ runs over
$u,d,s,c,b$ if we are at the $Z$ pole.
Write
$$
  F_{9,\hky}(x,\xb)\ =\ {\alpha_s\over\pi} g_{9,\hky}^{(1)}
  {(\xb-x)(x+\xb-1) \over 2(1-x)(1-\xb)}
  \sqrt{(1-x)(1-\xb)(x+\xb-1)-z(2-x-\xb)^2}\ \Im f,
\eqn\hkyfnine
$$
where
$$
   g_{9,\hky}^{(1)}\ =\ \Bigl( 2V_eA_e-(V_e^2+A_e^2)\Pe \Bigr)
   \ A_q |\chi(s)|^2
\eqn\hkygnineonedef
$$
and
$$
  \Im f\ \equiv\ \sum_{i={\rm flavors}} (2I_3^i) \, \Im f^i.
\eqn\imfdef
$$
For $4m_i^2 > s$, $\Im f^i = 0$.  (The top quark does not
contribute in the loop here.)
For $4m_i^2 < s$, let $z_i = m_i^2/s$.
Then there are two kinematical cases,
$$
\eqalign{
  \Im f^i\ &=\ {\pi \over 4(2-x-\xb)^2} \Biggl[
   4z_i \Bigl( \ln\bigl(\sqrt{x+\xb-1}+\sqrt{x+\xb-1-4z_i}\bigr)
          - \ln\bigl(1+\sqrt{1-4z_i}\bigr) \Bigr) \cr
&\qquad\qquad\qquad\qquad
   - \sqrt{1-{4z_i \over x+\xb-1}} + \sqrt{1-4z_i} \Biggr]\ ,
  \qquad  x+\xb > 1+4 z_i, \cr
  \Im f^i\ &=\ {\pi \over 4(2-x-\xb)^2} \Biggl[
   4z_i \Bigl( \ln\bigl(\sqrt{4z_i}\bigr)
          - \ln\bigl(1+\sqrt{1-4z_i}\bigr) \Bigr)
      + \sqrt{1-4z_i} \Biggr]\ , \qquad  x+\xb < 1+4 z_i. \cr
}\eqn\imfidef
$$
Setting $z=0$ we recover the results of ref~[\use\HKY].
Note that (\hkyfnine) is proportional to the axial coupling of both the
internal quark and the final-state quark. In particular, the
contributions of the $(u,d)$ and $(c,s)$ isospin doublets
in the final state cancel, up to the small effects of mass-splittings.
If the final-state flavor is not tagged,
one may therefore
keep only $b\bar{b}g$ final-state contributions to the triple-product
asymmetries.

In the limit of small internal quark mass, $z_i \to 0$,
$\Im f^i$ vanishes like $z_i$,
except for the small kinematical strip where $x+\xb < 1+4z_i$,
where $\Im f^i$ is $\Ord(1)$.
If we also neglect the external quark mass (set $z=0$),
then we get
$$
\eqalign{
  F_{9,\hky}(x,\xb)\ &\approx\
  \alpha_s g_{9,\hky}^{(1)} (2I_3^i) \, z_i \,
  \sqrt{ x+\xb-1 \over (1-x)(1-\xb)}
  {(\xb-x)\Bigl( (x+\xb-1)\ln(x+\xb-1) + 2-x-\xb \Bigr)
    \over 4(2-x-\xb)^2}\ , \cr
&\qquad\qquad x+\xb > 1+4z_i, \cr\cr
&\approx\
  \alpha_s g_{9,\hky}^{(1)} (2I_3^i)
  \sqrt{ x+\xb-1 \over (1-x)(1-\xb)}
  {(\xb-x)(x+\xb-1) \over 8(2-x-\xb)^2}\ , \cr
&\qquad\qquad x+\xb < 1+4z_i. \cr
}\eqn\hkyfninesmallz
$$

\vskip0.3truein
\par\noindent
{\bf 3. Electroweak $e^+e^- \to (\gamma^*,Z) \to q\qb g$
Contribution \hfill }
\par\noindent
We neglect all external quark masses in this contribution.
Denote the mass of the exchanged vector boson by $M_i$, $i=Z,W$,
and let $\xi_i\equiv M_i^2/s$.
At the $Z$ pole, $\xi_Z=1$, $\xi_W = 0.774$.
Write
$$
\eqalign{
  F_{9,\ew}(x,\xb)\ &=\ -{\alpha \over \sstw\cstw}
    \sqrt{ x+\xb-1 \over (1-x)(1-\xb) } \cr
&\qquad\times
 \sum_{i=Z,W} g_{9,\ew}^{(1),i}
 \biggl[  f_{Zq\qb}(\xi_i)
        + f_{g\qb}(\xi_i) \, \Theta(1-x-\xi_i)
        + f_{qg}(\xi_i)   \, \Theta(1-\xb-\xi_i)
 \biggr]\ , \cr
}\eqn\ewfnine
$$
where $\Theta(y)$ is the Heaviside step function,
$\Theta(y)=0$ for $y<0$, $\Theta(y)=1$ for $y>0$.
The function $f_{qg}$ represents the contribution from rescattering
in the $qg$ channel (the channel with momentum $k_q+k_g$ flowing
through it).  Since $(k_q+k_g)^2 > M_i^2$ is required for a nonzero
contribution in this channel, only the $W$ contributes to the $f_{qg}$
term, and likewise to the $f_{g\qb}$ term, which arises from
rescattering in the $g\qb$ channel.
Finally, $f_{Zq\qb}$ comes from the sum of the $Z$ channel
(carrying momentum $k_{e^+}+k_{e^-}$) and the $q\qb$ channel
contributions; both $Z$ and $W$ contribute here.

The couplings $g_{9,\ew}^{(1),i}$ are given by
$$
\eqalign{
  g_{9,\ew}^{(1),i}\ =\
& \Bigl( \bigl(V_q^{(i)}\bigr)^2 + \bigl(A_q^{(i)}\bigr)^2 \Bigr)
 \biggl[
   \Bigl( -2V_eA_e+(V_e^2+A_e^2)\Pe \Bigr) (V_q^2+A_q^2) |\chi(s)|^2
   \cr
&\qquad\qquad\qquad
      \ -\ 2Q(V_e\Pe - A_e)V_q \Re\chi(s)  \ +\ Q^2 \Pe
 \biggr], \cr
&\quad +\ \Bigl( -2V_eA_e+(V_e^2+A_e^2)\Pe \Bigr) \times
  4 V_q A_q V_q^{(i)} A_q^{(i)} \, |\chi(s)|^2\ , \cr
}\eqn\ewgnineone
$$
where the $Z$ vector and axial couplings are given by
$V_q^{(Z)}=V_q$ and $A_q^{(Z)}=A_q$, as listed in formula~(\vachidef),
and the corresponding expressions for $W$ exchange are
$$
   V_q^{(W)}\ =\ A_q^{(W)}\ =\ {-1\over2\sqrt{2}}\ctw,
   \qquad q=u,d,s,c,
\eqn\nonbVA
$$
and
$$
   V_b^{(W)}\ =\ A_b^{(W)}\ =\ 0.
\eqn\bVA
$$
In the above we neglect the small off-diagonal CKM matrix elements.
The special equation~(\bVA) is due to the fact that producing a
$\bar{b}$ pair after a $W$ exchange requires a $t\bar{t}$ to be present
in an intermediate state;
but below the $t\bar{t}$ threshold such a graph
cannot have an imaginary part.

The kinematic functions are
$$
\eqalign{
  f_{Zq\qb}(\xi)\ &=\
  - {3(1-\xb) \over \xi^2} \ell_3\biggl(\frac{x}{\xi}\biggr)
  + {3(1-x) \over \xi^2}   \ell_3\biggl(\frac{\xb}{\xi}\biggr) \cr
&\quad
  - (1-\xb)\biggl[ {(1-\xb) \over \xi(x+\xb-1)}+{2 \over \xi^2}\biggr]
     \ell_2\biggl(\frac{x}{\xi}\biggr)
  + (1-x)\biggl[ {(1-x) \over \xi(x+\xb-1)}+{2 \over \xi^2}\biggr]
     \ell_2\biggl(\frac{\xb}{\xi}\biggr) \cr
&\quad
  + {(1-\xb)^2 \over (x+\xb-1)(x+\xb-1+\xi)}
     \ell_2\biggl(\frac{1-\xb}{x+\xb-1+\xi}\biggr) \cr
&\quad
  - {(1-x)^2 \over (x+\xb-1)(x+\xb-1+\xi)}
     \ell_2\biggl(\frac{1-x}{x+\xb-1+\xi}\biggr), \cr
}\eqn\fzqq
$$
$$
\eqalign{
  f_{g\qb}(\xi)\ &=\
 {3(1-\xb)(1-x-\xi)^4 \over \xi^2}
    \ell_3\biggl(\frac{x(1-x-\xi)}{\xi}\biggr)
+ {(2-\xi)(1-\xb)(1-x-\xi)^3 \over \xi^2}
    \ell_2\biggl(\frac{x(1-x-\xi)}{\xi}\biggr) \cr
&\quad
+ {(1-\xb)(1-x-\xi)^2 \over x+\xb-1}
    \ell_1\biggl(\frac{x(1-x-\xi)}{\xi}\biggr) \cr
&\quad
- {(x+\xi)(x+\xb-1)(1-x-\xi)^2 \over \xi(1-x)}
    \ell_1\biggl(\frac{(1-x-\xi)(x+\xb-1)}{\xi}\biggr) \cr
&\quad
+ {(x+\xb-1+\xi)^2 \over (1-\xb)(x+\xb-1)}
    \ln\biggl(\frac{\xi+(1-x-\xi)(x+\xb-1)}{(x+\xi)(1-x)}\biggr) \cr
&\quad
- (1-x-\xi) \biggl[
      {(1-x)(1-\xb) \over \xi}
    - {\xi \over x+\xb-1}
    + {x(1-x-\xi) \over (1-x)^2}
    - {\xi(1+\xb)+2\xb \over 2(1-x)}
    - {3\over2} (1-\xb) \biggr]\ , \cr
}\eqn\fgq
$$
$$
\eqalign{
  f_{qg}(\xi)\ &=\ -f_{g\qb}(\xi)|_{x\lr\xb}, \cr
}\eqn\fqg
$$
where
$$
\eqalign{
  \ell_1(y)\ &=\ {\ln(1+y)-y \over y^2}\ , \cr
  \ell_2(y)\ &=\ {\ln(1+y)-y+y^2/2 \over y^3}\ , \cr
  \ell_3(y)\ &=\ {\ln(1+y)-y+y^2/2-y^3/3 \over y^4}\ . \cr}
$$
Note that the $\ell_i(y)$ are nonsingular as $y\to0$.

\vskip0.3truein
\par\noindent
{\bf 4. Non-standard-model ``$B$'' Gauge Boson Exchange
 Contribution \hfill }
\par\noindent
As in the electroweak case, we neglect all external quark masses.
Denote the mass of the exchanged vector boson by $M_B$, and
following the conventions of~[\use\CM,\use\BD], let it
couple vectorially to quarks with strength $\alpha_B/9$
(since quarks have baryon number $1/3$).
Let $\xi_B\equiv M_B^2/M_Z^2$.
Then the electroweak formulas can be modified to give,
$$
\eqalign{
  F_{9,B}(x,\xb)\ &=\ -{\alpha_B \over 9}
   \sqrt{ x+\xb-1 \over (1-x)(1-\xb) } \cr
&\times
  g_{9,B}^{(1)}
 \biggl[  f_{Zq\qb}(\xi_B)
        + f_{g\qb}(\xi_B) \, \Theta(1-x-\xi_B)
        + f_{qg}(\xi_B)   \, \Theta(1-\xb-\xi_B)
 \biggr]\ . \cr
}\eqn\Bfnine
$$
The couplings $g_{9,B}^{(1)}$ are now given by
$$
\eqalign{
  g_{9,B}^{(1)}\ &=\
 \Bigl( -2V_eA_e+(V_e^2+A_e^2)\Pe \Bigr) (V_q^2+A_q^2) |\chi(s)|^2
      \ -\ 2Q(V_e\Pe - A_e)V_q \Re\chi(s)  \ +\ Q^2 \Pe\ , \cr
}\eqn\Bgnineone
$$
and the functions $f_{Zq\qb}(\xi)$, $f_{g\qb}(\xi)$, $f_{qg}(\xi)$ are
exactly as in~(\fzqq)--(\fqg).

\section{Numerical Results}
\tagsection\NumericalSection

In this section, we present numerical results for the sizes of the
triple product correlations, and their dependence on the center-of-mass
energy and the three-jet cut. These results are
obtained by identifying jet momenta with parton momenta.
Several different ``event handedness''  correlations can be constructed
for the process we are considering.
Here we discuss the different contributions of section~3 to the
triple product correlation introduced in (\use\triplecorr),
$\langle {\bf \hat{k}_e \cdot (k_1 \times k_2) } \rangle$, and the
expectation value of
the normal angle, as signed by the two fastest jets,
$$
  \cos\theta_n\ =\
  { {\bf \hat{k}_e} \cdot ({\bf k_1}\times {\bf k_2})
    \over |{\bf k_1}\times {\bf k_2}| }\ ,
\eqn\thnormalfastdef
$$
where ${\bf k_i}$ are the energy-ordered momentum vectors
($E_1>E_2>E_3$).
As the two observables are qualitatively similar,
we only give numerical results for
$\costetn$.
Other variations, such as
${\bf \hat{k}_e} \cdot ({\bf k_1}\times {\bf k_2})$/
$(|{\bf k_1}|\ |{\bf k_2}|)$,
lead to similar or smaller signals.

The normal angle $\theta_n$, defined by the energy ordering, is
equal to the normal angle $\thnormal$, defined by the quark and
anti-quark, up to a sign
$$
\eta\ =\ \hbox{sign}\Bigl( (x-\xb)(x-x_g)(\xb-x_g) \Bigr),
\eqn\etadef
$$
where $x_g = 2-x-\xb$ is the gluon energy fraction.

Performing the angular integrals, the $\costh$ expectation value
is expressed in terms of $F_9$ and $F_1$ by
$$
\langle \cos\theta_n \rangle
\ =\ {1\over4}
{ \int_{D_c} dx\,d\xb\, \eta \, F_9(x,\xb)
\over
  \int_{D_c} dx\,d\xb\, F_1(x,\xb) }\ ,
\eqn\costhformula
$$
where $D_c$ is the domain in $(x,\xb)$ after making some
kind of a three-jet cut.
Similarly,
$$
\langle {\bf \hat{k}_e \cdot (k_1 \times k_2) } \rangle
\ =\ {s\over8}
{ \int_{D_c} dx\,d\xb\, \eta \sqrt{(1-x)(1-\xb)(x+\xb-1)-z(2-x-\xb)^2}
  \, F_9(x,\xb)
\over
  \int_{D_c} dx\,d\xb\, F_1(x,\xb) }\ .
\eqn\tripleprodformula
$$
On the $Z$ pole, the asymmetries~(\costhformula)
and~(\tripleprodformula)
 are proportional to the $Z$ polarization,
$$
  P_Z\ =\ {\Pe-A_{LR}^{(e)} \over 1-\Pe A_{LR}^{(e)} },
\eqn\pzperelation
$$
where $A_{LR}^{(e)} = 2V_eA_e/(V_e^2+A_e^2)$.
For example, at $s=M_Z^2$,
$$
\eqalign{
&\langle {\bf \hat{k}_e \cdot (k_1 \times k_2) } \rangle_\qcd \cr
&\ =\ P_Z {\alpha_s\over\pi} {M_Z^2\over8}
{ \int_{D_c} dx\,d\xb\, \eta
  \Bigl((1-x)(1-\xb)(x+\xb-1)-z(2-x-\xb)^2\Bigr)
  \, \Bigl( V_q^2 f_\qcd^v + A_q^2 f_\qcd^a \Bigr)
\over
  \Bigl( V_q^2 + A_q^2 \Bigr)
  \int_{D_c} dx\,d\xb\, F_1^{(0),v}}\ . \cr}
\eqn\tripleprodZFKSS
$$

Through most of our analysis, we use the standard
cut $y_{ij} \geq \ycut$,
where $y_{ij} \equiv\ (k_i+k_j)^2/M_Z^2$. Then, for massless quarks
$D_c$ is given by
$$
  D_c:\qquad x\leq 1-\ycut, \quad \xb\leq 1-\ycut,
    \quad x+\xb \geq 1+\ycut.
$$
With this cut, and if one neglects quark masses,
the leading order contribution to $F_1$
can be integrated analytically,
$$
\eqalign{
I_D(\ycut)\ &\equiv\ \int_{D_c} dx\,d\xb\, F_1^{(0),v}(x,\xb)
\ =\ \int_{D_c} dx\,d\xb\, {x^2+\xb^2\over2(1-x)(1-\xb)} \cr
\ &=\ \ln^2\left({1-\ycut\over\ycut}\right)
\ -\ {3\over2}(1-2\ycut)\ln\left({1-2\ycut\over\ycut}\right)
\ +\ 2\,\Li_2\left({\ycut\over1-\ycut}\right) \cr
&\qquad\qquad\qquad\ -\ {\pi^2\over6}\ + {5\over4}\ -\ 3\ycut
\ -\ {9\over4}\ycut^2. \cr}
\eqn\denomintegral
$$
We evaluate the remaining $x$, $\xb$ integrals
numerically.
As input for our
results we use $\alpha_s(M_Z)=0.116$, $m_b=4.5$~GeV,
$\sin^2\theta_W=0.232$, $M_Z=91.17$~GeV and $M_W=80.1$~GeV.
We always assume complete right-handed electron polarization:
$P_e=+1$.   It is easy to scale the results to
other values of $P_e$ at the $Z$ pole (using equation~(\use\pzperelation))
and well below the $Z$ pole
(where the observables are directly proportional to $P_e$).

At the $Z$-pole and below it,
the largest standard model effects arise from the
QCD contribution of section~3.1,
which is dominated (for $\sqrt{s}>2m_b$) by
the $b\bar{b}g$ final state. The $F_{9,\qcd}$ contribution to
$\costetn$
is shown in figure~7 as a function of the center-of-mass energy
$\sqrt{s}$, with $\ycut=0.04$, for $b$ production only.
If the final state is not flavor
tagged, then one should average over final state flavors, and the
result would be diluted by the fraction of hadronic events containing
$b$ quarks.  (At the $Z$, the $b$
fraction is $R_b = \Gamma_b/\Gamma_{{\rm hadron}}\approx 0.22$.)

As expected, the signal decreases with increasing $s$, roughly as $m_b^2/s$.
A further suppression
arises because the vector and axial components of the signal have opposite
signs. The dotted line shows the vector component of the result,
obtained by setting the $Zb\bar{b}$ axial coupling to zero. At small energies,
the signal is dominated by the vector component, which is
positive. At
larger energies, the axial component sets in with an opposite sign, and
exactly cancels the vector component just below the $Z$ mass. At
energies above the $Z$ mass, the signal is dominated by the axial
component. Notice that for center of mass energies below 30~GeV, one
would have to increase $\ycut$ in order to effectively cut soft gluons,
since $y_{ij}$ is always larger than $m_b^2$/$s$.

Near the $Z$ pole, the combined $W$ and $Z$ exchange contribution to
$\costetn$ (from $F_{9,\ew}$),
is about $+30\%$ of the QCD contribution in an untagged sample.
This is partly due
to the $n_f$ enhancement of the electroweak contribution; all final
state flavors contribute to the asymmetry (except the $b$ in $W$
exchange), whereas, practically, only $b$ quarks
contribute in the QCD case.
The resulting $\costetn$, assuming $b$ quarks are not tagged, and
summing over all flavors contributing to the asymmetry in the
electroweak contributions, is shown in figure~8 for $\sqrt{s}$ of
70--200~GeV.
For high energies,
the $W$ exchange contribution becomes dominant.
However, recall that we have neglected the contributions with
$\gamma\gamma$ and $\gamma Z$ intermediate states (figure~4);
and above the $W$-pair threshold, additional diagrams with
$WW$ intermediate states will contribute as well.
At these energies, pure hadronic decays of real $W$ pairs will form a
large ``background'' to the measurement.
(Triple product correlations in $e^+e^-\to W^+W^-$ via electroweak
rescattering are discussed in ref.~[\use\BILAL].)
Here we only plot the center-of-mass energy
dependence of the particular contributions we studied.

The second type of QCD rescattering, via the $Zg^*g$ effective vertex
discussed in section~3.2, gives rise to
asymmetries that are two to three orders of magnitude smaller than the
contributions mentioned above. At the $Z$, with $\ycut = 0.04$ for
example, $\costetn_{\hky} = -0.95 \times 10^{-8}$. We therefore neglect
this contribution in the remainder of the section.

As can be seen, the standard model prediction for the asymmetry at the
$Z$ is tiny. One might wonder whether it would change significantly
with the choice of
the three-jet cut. The T$_{\rm N}$-odd correlations should be
small both for large values of $\ycut$, which imply an almost symmetric
three-jet event,
and for small values of $\ycut$, which include soft or
collinear regions, where the event is two-jet like. Indeed, at the $Z$,
the QCD contribution to $\costetn$ peaks slightly below $\ycut=0.02$,
and the electroweak contributions peak near $\ycut=0.04$.

However, the relevant quantities to consider in order to determine the
optimal cut are not the values of the observables themselves, but rather the
corresponding signal-to-noise ratios, which describe the
statistical significance of a measurement.
The noise comes from root-mean-square fluctuations in the
T$_{{\rm N}}$-even cross-section. At lowest-order in $\alpha_s$
(tree-level), and neglecting quark masses, these can be calculated
analytically. We find,
$$
\Delta  \cos\theta_n\
\equiv\
\sqrt{\langle \cos^2\theta_n\rangle-\langle \cos\theta_n
\rangle^2}
\
\approx\ \sqrt{\langle \cos^2\theta_n\rangle}
\ =\ \sqrt{{3\over 10}},
\eqn\deltacos
$$
and
$$
\eqalign{
\Delta {\bf \hat{k}_e \cdot (k_1 \times k_2) }\ &\approx\
\sqrt{ \langle ({\bf \hat{k}_e \cdot (k_1 \times k_2) })^2 \rangle }
\cr \ &=\ {s\over 20}\sqrt{{3 (1-3\ycut)^2
(2-3\ycut+4\ycut^2+\ycut^3)
\over 2I_D(\ycut)}}.\cr}
\eqn\deltatriple
$$
The corresponding signal-to-noise ($S/N$) ratio for $\costetn$,
is then given by
$$
S/N(\cos\theta_n)\equiv {|\langle \cos\theta_n\rangle|
\over \Delta \cos\theta_n} \sqrt{N_{3-{\rm jet}}},
$$
and similarly for the triple product, where $N_{3-{\rm jet}}$ is the
number of three-jet events in the data sample.

The signal-to-noise ratios for $\costetn$, from
QCD and electroweak rescattering at the $Z$, are shown in figure~9 as
functions of $\ycut$.
In the range shown, the ratios increase monotonically as $\ycut$ decreases.
They eventually start to fall off as expected, but this happens for
very low values of the cut, where the perturbative calculation cannot
be trusted.
It is easy to understand why the $S/N$ ratio peaks at a lower $\ycut$
as does $\costetn$. The $F_9$ integrals are
finite for small $\ycut$, whereas the $F_1$ integrals diverge as double
logarithms for massless quarks, or logarithmically for $b$ production.
(The $b$ mass cuts off the collinear divergences). The signal to noise
ratio is proportional to the $F_9$ integral divided by the {\it square
root} of the $F_1$ integral, and so falls off more slowly than
$\costetn$ for small $\ycut$.
The QCD signal-to-noise ratio
continues to grow down to $\ycut=0.003$,
even though the $F_9$ integral is finite in the soft gluon region.
This suggests that it receives large contributions in regions where two of
the jets are close to collinear. We will return to this point later.
We note that replacing $\ycut$ by a cut on the smallest jet energy
leads to a smaller asymmetry.

We now turn to the effects of the hypothetical $B$-boson of
section~(3.4).
The contribution of the $B$ boson
to $\costetn$ at the $Z$, is given by the solid line in figure~10,
as a function of $\xi = m_B^2/M_Z^2$.
Here we take the $B$ coupling to be $\alpha_B$ = 0.2/9.
Up to overall factors which involve the couplings,
the $Z$ and $W$ contributions can be read off this plot, at
$\xi = \xi_Z = 1$ and $\xi = \xi_W = 0.774$ respectively.
The asymmetry is most sensitive to the $B$ boson if its mass is around
25--30~GeV. But even for a mass in this range the signal is probably too
small to be observed ($\costetn\sim 3\times10^{-5}$ or less).

The asymmetries~(\costhformula) and~(\tripleprodformula) involve
integrating $F_9$ with the sign $\eta$. Kinematic
regions with different energy orderings contribute with different
signs and potentially
cancel each other. Such cancellations would be avoided if the
gluon jet could be identified, so that the asymmetries could be defined
according to the energy ordering of the $q$, $\bar{q}$ jets only.
(For example, taking $k_1$, $k_2$ in~(\tripleprod)
to be the $q$, $\qb$ momenta, with
$E_1 > E_2$, so that $\eta = \hbox{sign}(x-\xb)$ in~(\etadef).)
This leads to little
improvement for the QCD contribution: $\costetn$ hardly changes, and
$\langle {\bf \hat{k}_e \cdot (k_1 \times k_2) } \rangle$ increases by
a factor of two to three, depending on the cut.
The effect is more significant for the electroweak and hypothetical
$B$-boson contributions to the  asymmetries, which
increase by a factor of around six. The $B$-boson contribution to
$\costetn$ at the $Z$, assuming the gluon jet is identified, is given
by the dashed line in figure~10.
Without gluon identification, the maximum signal-to-noise ratio
obtained is 0.17$\sqrt{{\cal L}/{\rm fb}^{-1}}$,
where $\cal L$ is the integrated luminosity
in inverse femtobarns.
If the gluon is identified with efficiency
$\epsilon_g$, this becomes
$0.6\sqrt{\epsilon_g{\cal L}/{\rm fb}^{-1}}$,
assuming $100\%$ purity.

Another way of enhancing the asymmetry is to use an ``optimized''
observable~[\use\OPT],
i.e., an observable that maximizes the signal-to-noise
ratio.
If the identity of the particles making up the jets is not
known, then the optimized observable is given by:
$$
{\widetilde O(x_1,x_2)}\ =\
{\cos\theta_n\over ({3\over 2}-{1\over 2}\cos^2\theta_n)}
\ \times\
{\sum_{p=1}^6 \eta_p\,F_9(x_p,\xb_p)
\over
\sum_{p=1}^6 F_1(x_p,\xb_p) }\ ,
\eqn\optobs
$$
where $x_1$, $x_2$ are the energy fractions of the highest energy and
intermediate energy jet respectively,
$p$ sums over the six different ways of assigning $x_1$, $x_2$, and
$2-x_1-x_2$ to $x$, $\xb$ and $x_g=2-x-\xb$, and $\eta_p$ is $\eta$
of~(\etadef), evaluated with $x=x_p$, $\xb=\xb_p$.

The optimized observable signal-to-noise ratios are only
20--30$\%$ bigger than the $\costetn$ signal-to-noise ratios
for the $W$, $Z$ and $B$-boson exchange contributions. This holds
whether or not the gluon is identified.
The same enhancement occurs for the QCD contribution, with no gluon
identification. If the gluon is identified, the enhancement is much
bigger. As mentioned above, the QCD asymmetries receive large
contributions from regions where two of the jets are close to
collinear. An ``upper limit'' estimation of the signal-to-noise that can
be produced by the QCD contribution at the
$Z$ is obtained by studying the optimized observable, assuming gluon
identification, and replacing $\ycut$ by a cut on the jet
energies: $E_i \geq E_{min}$. For $E_{min} = 5-10$~GeV we find,
for $b$ production only,
$$
{ \langle {\widetilde O}\rangle \over
 \langle {\widetilde O^2}\rangle^{1/2} }
\ =\ (1.5 - 1.9)\times 10^{-4}\ \ \ \
{\rm and}
\ \ \ \
{ \costetn\over
 \langle \cos^2\theta_n\rangle^{1/2} }
\ =\ (0.6 - 1.0)\times 10^{-4}\ ,
$$
giving signal-to-noise ratios of about
$$
{ S/N(\widetilde O})
\ =\ 0.33\sqrt{\epsilon_g\epsilon_b}\,\sqrt{{\cal L}/{\rm fb}^{-1}}\ \ \
{\rm and}
\ \ \ \
 S/N(\cos\theta_n)
 \ =\ 0.15\sqrt{\epsilon_g\epsilon_b}\,\sqrt{{\cal L}
/{\rm fb}^{-1}} \ ,
$$
where $\epsilon_g$, $\epsilon_b$ are the gluon identification and
$b$-tagging efficiencies.

Somewhat higher sensitivity to the QCD-induced asymmetry can be
achieved at low center-of-mass energies. For $\sqrt{s} = 30$~GeV,
the signal-to-noise ratio for $\costetn$, with $\ycut=0.04$,
is $0.3\sqrt{\epsilon_b}\sqrt{{\cal L}/{\rm fb}^{-1}}$, assuming
that the $b$ is tagged with efficiency $\epsilon_b$ and 100$\%$ purity.
With gluon identification, this result becomes
$0.4{\sqrt{\epsilon_b\epsilon_g}}\sqrt{{\cal L}/{\rm fb}^{-1}}$.
If $\ycut$ is replaced by $E_i\geq 2$~GeV, one finds
$S/N(\cos\theta_n) = 0.55\sqrt{\epsilon_b\epsilon_g}
\sqrt{{\cal L}/{\rm fb}^{-1}}$,
and
$S/N({\widetilde O}) = 0.7\sqrt{\epsilon_b\epsilon_g}
\sqrt{{\cal L}/{\rm fb}^{-1}}$.
In any case, integrated luminosities in at least the tens of
inverse-femtobarn range will be required for measurements of
these standard model contributions to be statistically significant.


\section{Conclusions}

Beam polarization, or the natural $Z$ polarization, can be used
to construct ``event handedness'' correlations in
$e^+e^- \to$~3-jets that are directly sensitive to
rescattering effects.
In this paper we have identified and calculated the dominant
standard model contributions to several such correlations in $e^+e^-$
annihilation at or below the $Z$ pole.

QCD rescattering of massless quarks does not produce any
``event handedness'' correlations at one loop in the purely
time-like kinematics of $e^+e^-$ annihilation through a single
gauge boson. The dominant standard model contributions to
these correlations are therefore produced by QCD rescattering of massive
quarks, which is suppressed by $m_b^2/M_Z^2$ at the $Z$ resonance,
and by electroweak rescattering, via $W$ and $Z$ exchange loops.
We have presented analytic results for the different contributions.
We have studied the dependence of the resulting asymmetries
on different kinematic variables of the process considered,
including the center-of-mass energy and the three-jet cut.
Due to various cancellations, the standard model does not generate
large effects; even for ``optimized observables'' the signal-to-noise
ratios are quite small.

Thus, a measurement of event handedness correlations may serve as a
probe of physics beyond the standard model and/or nonperturbative
effects in jet physics.

We have investigated the asymmetry generated in quark rescattering
through the exchange of a hypothetical gauge boson, coupling to baryon
number only. The effects are the largest, but would still be difficult to
observe, if the mass of this boson lies in the range of 25--30~GeV.

\vskip .2 cm
\noindent{\bf Note added}
\vskip .1 cm

The SLD collaboration has recently placed an experimental upper bound
on the magnitude of $\langle\cos\theta_n\rangle$ at the $Z$ pole,
obtaining 95\% C.L. limits of $-0.022 < \beta < 0.039$, where
$\beta={16\over9}{1\over P_Z}\,\langle\cos\theta_n\rangle$~[\use\SLDbound].

\vskip .2 cm
\noindent{\bf Acknowledgements}
\vskip .1 cm

We thank M.~Peskin for many
useful discussions, and for his careful reading of the paper.
We also thank J.D.~Bjorken and D.~Atwood for useful discussions,
and T.~Maruyama and P.~Burrows for suggesting
this work, for continual encouragement and for their comments on the
manuscript.
A.B. would also like to thank W.~Bernreuther for many
clarifying discussions.

\vfill\eject


\appendix{Vanishing of Event Handedness Correlations in Massless QCD}
\tagappendix\vanishproof

In this appendix we show that the one-loop QCD contribution to
triple-product correlations in $e^+e^-$
annihilation through a single gauge-boson vanishes, unless some of the
partons propagating around the loop are massive.
This argument generalizes to $n$-parton final states a previous argument by
K\"{o}rner and Schuler~[\use\KoernerSchuler] for the three-parton case,
$e^+e^- \to q\qb g$.

A nonzero triple-product correlation is produced by terms in the
differential cross-section that are proportional to
the Levi-Civita tensor $\varepsilon_{\mu\nu\sigma\rho}$
contracted with four of the five momentum vectors in the problem.
The contraction must be multiplied by the imaginary
part of some loop integral, in order to contribute to the
differential cross-section.

The loop integrals may be defined by analytic continuation from
the unphysical, Euclidean region, to the physical region.
Denote the external kinematic invariants by $s_{ij} = (k_i+k_j)^2$,
with $k_i$ the momentum of the $i$-th particle. All the $s_{ij}$
are negative in the Euclidean region,
and all loop integrals are manifestly real there.
Upon going into the physical region, some of the invariants
may change sign, and the integrals may develop
imaginary parts. However, the dependence of the integrals
on the kinematic invariants is through analytic functions of
dimensionless ratios, of the form
$f({-s_{ij} \over -s_{kl}})$.
The kinematics of $e^+e^-$ annihilation through a single gauge
boson is purely timelike --- all the invariants appearing in the loop
integrals are positive in the physical region. Therefore, ratios of
invariants do not change sign upon going from the Euclidean region to
the physical region. As a result, the integrals do not develop any
imaginary parts in the physical region, and the loop amplitude has no
absorptive part.

This ceases to be true if some particle propagating around the loop
has a non-zero mass $M$. In this case the
dimensionless functions appearing in the amplitude are of the
form
$f({-s_{ij} \over -s_{jk}},{M^2\over-s_{ij}})$.
Since $M^2$ is positive in both the Euclidean region and the physical
region, the ratios ${M^2\over-s_{ij}}$ flip sign as one goes from
Euclidean to physical, and imaginary parts are now permitted.

In the above, we ignored one source of a ``mass scale'', which is
present in perturbation theory even when all the particles are massless,
namely, the renormalization scale $\mu$.
The renormalized one-loop amplitude
for producing $n$ final state partons,
$\Aloop_n$, can be written as a sum of two pieces:
an infrared-divergent piece, and a leftover finite piece.
(This separation has some arbitrariness associated with it.)
The finite piece of $\Aloop_n$ depends only
on the kinematic invariants $s_{ij}$ and on particle masses, and
as we saw above, it has no absorptive part in the purely time-like
kinematics of the process we are considering.
In contrast, the infrared-divergent piece of
$\Aloop_n$ contains
logarithms of the form $\ln(\mu^2/(-s_{ij}))$,
which may and do develop imaginary parts in physical
regions, including the fully time-like region.
But, as we now show, these
imaginary parts do not contribute to the cross-section at one loop.

When interfered with $\Atree_n$,
the infrared-divergent piece of $\Aloop_n$
cancels soft and collinear
phase-space integrations of the tree-level cross-section
$|\Atree_{n+1}|^2$ for producing $n+1$ partons,
where one of the partons is unobserved.
Its form can therefore be inferred from the soft and collinear
structure of the cross-section for producing $n+1$ partons.
It can be written as a sum of terms, where each term is given
by a corresponding term in the tree amplitude $\Atree_n$,
multiplied by a factor that depends on a single
invariant $s_{ij}$ (see for example~[\use\KUNSZT]). This factor is
universal --- it only depends on the identity of partons $i$ and $j$,
i.e., on whether they are quarks or gluons.

The strongest singularities come from overlapping soft and collinear
regions, and can be written, using dimensional
regularization with $D = 4-2\epsilon$, as~[\use\KUNSZT]
$$
{\cal A}^{{\rm 1-loop, soft}}_{c_1\cdots c_n}\ =  \ \gamma
\sum_{i<j}\, S_{ij}\,
t^a_{c_ic_i^\prime}
t^a_{c_jc_j^\prime}\,
\Atree_{c_1\cdots c_i^\prime \cdots c_j^\prime\cdots c_n}\ ,
\eqn\softsing
$$
where $c_i$ is the color index of the $i$-th parton,
$t^a$ are the $SU(N)$ generators, $\gamma$ is a real constant, and
$$
S_{ij}\ =\ {1\over\epsilon^2}
\, \left({\mu^2\over -s_{ij} }\right)^\epsilon\ .
\eqn\softfactor
$$
When expanded around $\epsilon = 0$, the factor $S_{ij}$ contains
the logarithm $\ln(\mu^2/(-s_{ij}))$ which develops an imaginary part in
the physical region.

But for each term that contains $S_{ij}$ in the
interference $\Aloop_n\,{\Atree_n}^*$, there corresponds an identical term in
${\Aloop_n}^*\,\Atree_n$, in which $S_{ij}$ is replaced by $S_{ij}^*$. The
imaginary part of $S_{ij}$ therefore drops out in the cross-section.
Specifically, the interference $\Aloop_n\,{\Atree_n}^*$ contains the
term
$$
S_{ij}\, t^a_{c_ic_i^\prime} t^a_{c_jc_j^\prime}\,
\Atree_{c_1\cdots c_i^\prime \cdots c_j^\prime\cdots c_n}\,
\left(\Atree_{c_1\cdots c_n}\right)^*\ ,
\eqn\interfera
$$
while the interference $\Atree_n{\Aloop_n}^*$ contains the
term
$$
\Atree_{c_1\cdots c_n}
\left(S_{ij}\, t^a_{c_ic_i^\prime} t^a_{c_jc_j^\prime}\,
\Atree_{c_1\cdots c_i^\prime \cdots c_j^\prime\cdots c_n}\right)^*
\ = \
S_{ij}^*\, t^a_{c_ic_i^\prime} t^a_{c_jc_j^\prime}
\left(\Atree_{c_1\cdots c_n}\right)^*
\Atree_{c_1\cdots c_i^\prime \cdots c_j^\prime\cdots c_n}\ ,
\eqn\interferb
$$
where in the last line we used the hermiticity of $t^a$ to exchange
$c_i \lr c_i^\prime$, $c_j \lr c_j^\prime$. Thus, only the real part
of $S_{ij}$ contributes in the sum of~(\interfera) and~(\interferb).

This argument can be repeated for the hard collinear singularities,
which have the simpler form
$$
{\cal A}^{{\rm 1-loop, coll}}_{c_1\cdots c_n}\ =  \
\sum_{i<j}\,\gamma_{ij}\,{1\over\epsilon}
\, \left({\mu^2\over -s_{ij} }\right)^\epsilon\,
\Atree_{c_1\cdots c_n}\ ,
\eqn\collsing
$$
where again
$c_1\cdots c_n$ are the color indices of the partons and
$\gamma_{ij}$ are real constants that depend on the identity of partons
$i$ and $j$.

Notice that the argument does not rely on the fact
that~(\interfera) and (\interferb) contain the interference of two tree
amplitudes. The amplitudes appearing on the two sides of the
interference could in principle be different --- the crucial point is
that the singular factor can appear on
the two sides of the interference, multiplying the same structure.
It appears possible to generalize this argument beyond one loop, but we
have not yet done so.

\vfill\eject

\appendix{Kinematic Functions for QCD Massive Quark Contributions}
\tagappendix\FKSSFormulaeAppendix

In this appendix we give the kinematic functions that contribute to the
tree-level denominator of the expectation value~(\triplecorr),
keeping the quark mass nonzero~[\use\stav], followed by the
functions appearing in the one-loop QCD $(\gamma^*,Z)\to b\bar{b}g$
contribution to $F_9$.
In both cases we define $z=m_q^2/s$ with $m_q$ the external quark mass.

The tree-level $e^+e^- \to q\bar{q}g$ kinematic functions appearing
in~(\use\fiapprox) are
$$
\eqalign{
  F_1^{(0),v}\ &=\ { x^2+\xb^2 \over 2(1-x)(1-\xb) }
  + { z \bigl( 2x\xb(x+\xb)-3x^2-3\xb^2+8(x+\xb-x\xb)-6 \bigr)
      \over (1-x)^2(1-\xb)^2 } \cr
&\quad
  - {2z^2(2-x-\xb)^2 \over (1-x)^2(1-\xb)^2 }, \cr
  F_1^{(0),a}\ &=\ { x^2+\xb^2 \over 2(1-x)(1-\xb) } \cr
&\quad
  + { z \bigl( -(x+\xb)^2(x+\xb-x\xb) + 8(x^2(1-\xb)+\xb^2(1-x))
               + 24x\xb - 12(x+\xb) + 4 \bigr)
      \over (1-x)^2(1-\xb)^2 } \cr
&\quad
  + {4z^2(2-x-\xb)^2 \over (1-x)^2(1-\xb)^2 }, \cr
  F_2^{(0),v}-F_5^{(0),v}\ &=\
   { 2z \bigl( (1-x)(1-\xb)(x+\xb-1) - z(2-x-\xb)^2 \bigr)
      \over (1-x)^2(1-\xb)^2 }, \cr
  F_2^{(0),a}-F_5^{(0),a}\ &=\
   { z(2-x-\xb)^2 \over (1-x)^2(1-\xb)^2 }, \cr
}\eqn\massivetreefone
$$
with $z = m_q^2/s$.
For zero mass these expressions reduce to equations~(\use\masslesstreef).

Next we give the functions appearing in
equations~(\use\fkssfnine), (\use\ncdecomp) for the
QCD $e^+e^- \to (\gamma^*,Z)\to b\bar{b}g$ contribution to $F_9$.
We first decompose $f^{v,1}$, $f^{v,2}$, $f^{a,1}$ and $f^{a,2}$
into sums of imaginary parts of scalar integrals, multiplied by coefficient
functions,
$$
\eqalign{
  f^{v(a),1}\ &=\ d_{D=6}^{v(a)} \Im D_0^{D=6} \cr
&\quad   +c_{134}^{v(a),1}
\Im C_0(1,3,4) \ + c_{234}^{v(a),1} \Im C_0(2,3,4)\cr
&\quad        +b_{13}^{v(a),1} \Im B_0(1,3)
\ +b_{24}^{v(a),1}
\Im B_0(2,4) \ +b_{34}^{v(a),1}
\Im B_0(3,4) , \cr
  f^{v(a),2}\ &=\ \tilde{d}_{D=6}^{v(a)} \Im \tilde{D}_0^{D=6}\ +
 \tilde{d'}_{D=6}^{v(a)} \Im \tilde{D'}_0^{D=6} \cr
&\quad
+c_{134}^{v(a),2}
\Im C_0(1,3,4) \ + c_{234}^{v(a),2}\Im C_0(2,3,4)\cr
&\quad   +
\tilde{c}_{123}^{v(a)} \Im \tilde{C}_0(1,2,3)
\ +
\tilde{c}_{134}^{v(a)} \Im \tilde{C}_0(1,3,4) \ +
\tilde{c}_{234}^{v(a)} \Im \tilde{C}_0(2,3,4)\cr
&\quad        +b_{13}^{v(a),2}\Im B_0(1,3)
\ +b_{24}^{v(a),2}
\Im B_0(2,4) \ + b_{34}^{v(a),2}
\Im B_0(3,4)\cr
&\quad   + \tilde{b}_{24}^{v(a)}
\Im \tilde{B}_0(2,4).
 }\eqn\coeffdec
$$
Here $B_0$ stands generically for a bubble integral, $C_0$ for a
triangle integral, and $D_0^{D=6}$ for a ``$D=6$'' box
integral; the usual $D=4$ scalar box integral $D_0$
has been eliminated in favor of a linear combination of $D_0^{D=6}$ and
four $C_0$'s~[\use\OurIntegrals].

\vfill\eject

The explicit formulae for the imaginary parts of the integrals are
$$
\eqalign{
\Im D_0^{D=6} \ &=\ {-\pi \over s}
\biggl[ {\rho(1-2\xps)+
(1-\xb)\xps
\lrho (\xps)
 \over (\xps-\xms)(\xps-\xpt)(\xps-\xmt)} +
{\rho(1-2\xms)+
(1-\xb)\xms
\lrho (\xms)
\over (\xms-\xps)(\xms-\xpt)(\xms-\xmt)} \cr &\quad
+{\rho(1-2\xpt)+
(1-x)(1-\xpt)
\lrho(\xpt)
\over (\xpt-\xps)(\xpt-\xms)(\xpt-\xmt)}
 + {\rho(1-2\xmt)+
(1-x)(1-\xmt)
\lrho(\xmt)
\over (\xmt-\xps)(\xmt-\xms)(\xmt-\xpt)}
\biggr] , \cr
\Im C_0(1,3,4) \ &=\ {-\pi\over s \sqrt{\xb^2-4z}}
(\lrho(\xps)-\lrho(\xms)), \cr
\Im C_0(2,3,4) \ &=\ \Im C_0(1,3,4)|_{x\lr\xb}, \cr
\Im B_0(1,3) \ &=\  {\pi (1-\xb)\over (1-\xb+z)} ,\cr
\Im B_0(2,4) \ &=\  \Im B_0(1,3)|_{x\lr\xb} \cr
\Im B_0(3,4) \ &=\  \pi \sqrt{1-4z}, \cr
\Im \tilde{D}_0^{D=6} \ &=\
{\pi \over 2s((x+\xb-1)(1-x)(1-\xb)-z(2-x-\xb)^2)} \cr &\quad
\times \biggl\{ (x+\xb-1)(1-\xb)(1-2\rho')[2\ln(1-\xb)-4\ln(1-\rho-\rho')+
3\ln(1-\rho')-\ln(\rho')] \cr &\quad
+{(1-\xb)(\xb(x+\xb-1)-2z(x+\xb))\over \sqrt{\xb^2-4z}}[\lrho(\xps)
-\lrho(\xms)] \cr &\quad
+((x+\xb-1)(1-\xb)-2z(2-x-\xb))\biggl[\ln\biggl({z\over 1-\xb+z}\biggr)
-\ln\biggl({\rho\over 1-\rho}\biggr)+
\ln\biggl({\rho'\over 1-\rho'}\biggr)\biggr]\biggr\}, \cr
\Im \tilde{D'}_0^{D=6} \ &=\ \Im \tilde{D}_0^{D=6}|_{x\lr\xb}, \cr
\Im \tilde{C}_0(1,2,3) \ &=\  {\pi\over s(1-\xb)}
\ln \biggl({z\over 1-\xb+z}\biggr)             ,\cr
\Im \tilde{C}_0(1,3,4) \ &=\ \Im \tilde{C}_0(1,2,3)|_{x\lr\xb}  , \cr
\Im \tilde{C}_0(2,3,4) \ &=\  \ {\pi \over s(2-x-\xb)}
\biggl[ \ln \biggl( {\rho\over 1-\rho}\biggr)
- \ln \biggl( {\rho'\over 1-\rho'}\biggr) \biggr]
 , \cr
\Im \tilde{B}_0(2,4) \ &=\ \pi\sqrt{1-{4z\over x+\xb-1}},
}\eqn\imagint
$$
where
$$
\eqalign{
\rho \ &=\ {1-\sqrt{1-4z} \over 2}, \cr
\rho' \ &=\ {1\over 2}\biggl( 1-\sqrt{1-{4z\over x+\xb-1}}\biggr), \cr
x_{\pm}^s \ &=\ {\xb\pm \sqrt{\xb^2-4z} \over 2}, \cr
x_{\pm}^t \ &=\ 1-{x\mp \sqrt{x^2-4z} \over 2}, \cr
\lrho(y) \ &=\ \ln\biggl( {(1-y)(y-\rho) \over y(1-y-\rho)}\biggr). \cr
}\eqn\imagintdef
$$

The coefficient functions are
$$
\eqalign{
  d_{D=6}^{v} \ &=\
     {-z s (x-\xb)(2-x-\xb) \over (1-x)^2(1-\xb)^2},\cr
  c_{134}^{v,1} \ &=\
    \biggl[ -{\xb^2-4z \over (1-\xb)}
  -{\xb+2x\xb^2-3x\xb+2x-2\xb^2 \over (1-x)(1-\xb)}
  +{3x\xb -7x+2\xb^3-\xb-\xb^2+4 \over 2(\xb^2-4z)}
    \cr &\quad
  + {3\xb(1-\xb)(2(1-x-\xb)+\xb^2+x\xb) \over 2(\xb^2-4z)^2}
  \biggr]  \,{z s\over  2(1-x)} \, , \cr
  c_{234}^{v,1} \ &=\ -c_{134}^{v,1}|_{x\lr\xb}, \cr
  b_{13}^{v,1} \ &=\
    {-1 \over 32(1-x)(1+z-\xb)}
    \biggl[ {(\xb^2-4z)^3 \over (1-\xb)^2}
  - {(\xb^2-4z)^2(8+4\xb^2-9\xb) \over (1-\xb)^2} \cr &\quad
  + {(\xb^2-4z)(12\xb^4-51\xb^3+44+3x\xb^2-10x\xb-91\xb+92\xb^2+7x)
  \over 2(1-\xb)^2} \cr &\quad
  -{-45\xb^5+8\xb^6-50x\xb^3+16+88\xb^2+109x\xb^2-133\xb^3+106\xb^4
     +9x\xb^4+36x-104x\xb-38\xb \over 2(1-\xb)^2} \cr &\quad
  + {(2-\xb)(2\xb^6-3\xb^5+9x\xb^4-13\xb^4+54\xb^3-43x\xb^3-80\xb^2
     +78x\xb^2+60\xb-64x\xb-16+16x) \over 2(\xb^2-4z)(1-\xb)}
   \cr &\quad
   +{3\xb^2(2-\xb)^3(2(1-x-\xb)+\xb^2+x\xb) \over 2(\xb^2-4z)^2} \biggr]
, \cr
  b_{24}^{v,1} \ &=\ -b_{13}^{v,1}|_{x\lr\xb}, \cr
  b_{34}^{v,1} \ &=\
 {4x\xb^2-7x\xb-3\xb^2+2\xb+2\xb^3+2\xb^4+4-4x \over 8(1-x)(1-\xb)(\xb^2-4z)}
 \cr &\quad
 -{4x^2\xb-7x\xb-3x^2+2x+2x^3+2x^4+4-4\xb \over 8(1-x)(1-\xb)(x^2-4z)}
  \cr &\quad
  + {3\xb^2(2(1-x-\xb)+\xb^2+x\xb) \over 8(1-x)(\xb^2-4z)^2}
  - {3x^2(2(1-x-\xb)+x^2+x\xb) \over 8(1-\xb)(x^2-4z)^2} \cr &\quad
  - { (x-\xb)((x+\xb)(2(1-x-\xb)+x\xb)+2z(x+\xb+x\xb)+8z^2) \over
      2(1-x)(1-\xb)(x^2-4z)(\xb^2-4z)} \cr &\quad
  + {(x-\xb)(x+\xb-1) \over 4(1-x)(1-\xb) },
}
\eqn\rescoeffvone
$$
\vfil\eject
$$
\eqalign{
  d_{D=6}^{a}\ &=\
               {-z s(x-\xb)(3(1-x-\xb)+x^2+x\xb+\xb^2)
                  \over (1-x)^2(1-\xb)^2} ,\cr
   c_{134}^{a,1} \ &=\
    \biggl[{\xb^2-4z \over (1-\xb)}
   + {-9\xb^2+16\xb-10x\xb+5x\xb^2+7x-9+2x^2\xb-2x^2 \over 2(1-x)(1-\xb)}
 \cr &\quad -{-10x\xb+5x\xb^2+9x-13\xb^2+11\xb+6\xb^3-6
 \over 2(\xb^2-4z)} \cr &\quad
 +{3\xb(1-\xb)^2(2(1-x-\xb)+\xb^2+x\xb)\over 2(\xb^2-4z)^2} \biggr]
  \, {z s\over 2(1-x)} \, ,\cr
 c_{234}^{a,1} \ &=\ -c_{134}^{a,1}|_{x\lr\xb}, \cr
 b_{13}^{a,1} \ &=\
    {-1 \over 32(1-x)(1+z-\xb)}
    \biggl[ -{(\xb^2-4z)^3 \over (1-\xb)^2}
  +{(\xb^2-4z)^2(2x\xb+9+9\xb^2-18\xb)\over 2(1-\xb)^2} \cr &\quad
  -{(\xb^2-4z)(26+18\xb^4-73\xb^3-27x\xb^2+9x\xb^3+114\xb^2-91\xb
    -21x+43x\xb) \over 2(1-\xb)^2} \cr &\quad
  + \Bigl( -60x+382\xb^2-297x\xb^2+208x\xb+20\xb^6-479\xb^3+332\xb^4
  \cr &\quad +223x\xb^3
     -125\xb^5 -87x\xb^4
  +15x\xb^5+24-158\xb \Bigr) {1 \over 2(1-\xb)^2}\cr &\quad
  -\Bigl( 12\xb^6-67\xb^5+11x\xb^5-60x\xb^4+161\xb^4-212\xb^3 +131x\xb^3
     \cr &\quad -150x\xb^2+172\xb^2-84\xb+88x\xb+16-16x \Bigr)
{2-\xb \over 2(1-\xb)(\xb^2-4z)}
  \cr &\quad +{3\xb^2(1-\xb)(2-\xb)^3(2(1-x-\xb)+\xb^2+x\xb) \over
   2(\xb^2-4z)^2} \biggr] ,\cr
  b_{24}^{a,1} \ &=\ -b_{13}^{a,1}|_{x\lr\xb}, \cr
  b_{34}^{a,1} \ &=\
   -{23\xb^2+6x\xb^3-4-19x\xb^2-8\xb+7\xb^4-20\xb^3+13x\xb+4x \over
    8(1-x)(1-\xb)(\xb^2-4z)} \cr &\quad
    +{23x^2+6x^3\xb-4-19x^2\xb-8x+7x^4-20x^3+13x\xb+4\xb \over
    8(1-x)(1-\xb)(x^2-4z)} \cr &\quad
    +{3\xb^2(1-\xb)(2(1-x-\xb)+\xb^2+x\xb)\over 8(1-x)(\xb^2-4z)^2}
    -{3x^2(1-x)(2(1-x-\xb)+x^2+x\xb)\over 8(1-\xb)(x^2-4z)^2} \cr &\quad
    +{(x-\xb) \Bigl( -(x+\xb)(2(1-x-\xb)+x\xb)
       + 2z \bigl( x+\xb-2(x^2+\xb^2)-3x\xb \bigr) +8z^2 \Bigr)
     \over 2(1-x)(1-\xb)(x^2-4z)(\xb^2-4z)} \cr &\quad
    -{(x-\xb)(4(x+\xb)-1)\over 8(1-x)(1-\xb)},
}
\eqn\rescoeffaone
$$
\vfil\eject
$$
\eqalign{
\tilde{d}_{D=6}^{v}\ &=\
\biggl[ -(x+\xb-1)
  - {z (-3x+2x^2+\xb-\xb^2+x\xb)  \over (x+\xb-1)(1-\xb)}
  + {4z^2(2-x-\xb) \over (x+\xb-1)(1-\xb)}
  \biggr] \, { s \over 2(1-x)(1-\xb)}  \, , \cr
 \tilde{d'}_{D=6}^{v} \ &=\ -\tilde{d}_{D=6}^{v}|_{x\lr\xb}, \cr
 c_{134}^{v,2} \ &=\
 \biggl[ {(2-x-\xb)(\xb^2-4z)^2 \over (x+\xb-1)(1-\xb)}
         +{(\xb^2-4z)(-3+6\xb+2x-8\xb^2+3\xb^3+3x\xb^2-3x\xb)
    \over (x+\xb-1)(1-\xb)}\cr &\quad
+{(3(1-x-\xb)+\xb^3+x\xb)(3\xb-2+2x) \over (x+\xb-1)(1-\xb)}
 \cr &\quad
   -{-5\xb-6x\xb+2\xb^2-5x-5\xb^3+4+6\xb^4+3x\xb^2 \over 2(1-\xb)}
 \cr &\quad
   -{\xb(\xb^4+\xb^3-7\xb^2+3x\xb^2+10\xb-8x\xb-3+3x) \over (\xb^2-4z)}
 \cr &\quad
  -{3\xb^3(1-\xb)(2(1-x-\xb)+\xb^2+x\xb) \over 2(\xb^2-4z)^2}
\biggr]\,{ s\over 8 (1-x)} \, ,\cr
c_{234}^{v,2} \ &=\ -c_{134}^{v,2}|_{x\lr\xb}, \cr
\tilde{c}_{123}^{v}\ &=\ {-z s(x+4z) \over 2(x+\xb-1)(1-x)(1-\xb)}
,\cr
\tilde{c}_{134}^{v} \ &=\ -\tilde{c}_{123}^{v}|_{x\lr\xb}, \cr
\tilde{c}_{234}^{v} \ &=\ {-z s(x-\xb) \over (1-x)^2(1-\xb)^2}\
\biggl[{5(1-x-\xb)+x^2+3x\xb+\xb^2 \over 2-x-\xb}+
       {2z(2-x-\xb) \over x+\xb-1} \biggr] ,\cr
 b_{13}^{v,2} \ &=\
{1 \over 32 (1-x)(1+z-\xb)} \biggl[ {(\xb^2-4z)^3 \over (1-\xb)^2}
  -{(\xb^2-4z)^2(4\xb^2-9\xb+9) \over (1-\xb)^2} \cr &\quad
  +{(\xb^2-4z)(12\xb^4-51\xb^3+48+3x\xb^2-10x\xb-91\xb+92\xb^2+7x)
    \over 2(1-\xb)^2} \cr &\quad
  -{-45\xb^5+8\xb^6-50x\xb^3+24\xb^2+109x\xb^2-101\xb^3+100\xb^4
    +9x\xb^4+36x-104x\xb+18\xb \over 2(1-\xb)^2} \cr &\quad
  +{(2-\xb)(2\xb^6-3\xb^5+9x\xb^4-17\xb^4+74\xb^3-43x\xb^3-112\xb^2
   +78x\xb^2+76\xb-64x\xb-16+16x)\over 2(1-\xb)(\xb^2-4z)}
   \cr &\quad
 +{3\xb^2(2-\xb)^3(2(1-x-\xb)+\xb^2+x\xb) \over 2(\xb^2-4z)^2}\biggr], \cr
  b_{24}^{v,2} \ &=\ -b_{13}^{v,2}|_{x\lr\xb},
  }
\eqn\rescoeffvtwoone
$$
\vfil\eject
$$
\eqalign{
  b_{34}^{v,2} \ &=\
-{4x\xb^2-7x\xb-3\xb^2+2\xb+2\xb^3+2\xb^4+4-4x \over 8(1-x)(1-\xb)(\xb^2-4z)}
 \cr &\quad
 +{4x^2\xb-7x\xb-3x^2+2x+2x^3+2x^4+4-4\xb \over 8(1-x)(1-\xb)(x^2-4z)}
  \cr &\quad
  - {3\xb^2(2(1-x-\xb)+\xb^2+x\xb) \over 8(1-x)(\xb^2-4z)^2}
  + {3x^2(2(1-x-\xb)+x^2+x\xb) \over 8(1-\xb)(x^2-4z)^2}
 - {(x-\xb)(x+\xb-1)\over 4(1-x)(1-\xb)}
 \cr &\quad
+ \biggl( -2(x^4+\xb^4)+(x^3+\xb^3)(x\xb+10)-(x^2+\xb^2)(13x\xb+16)
           +(x+\xb)(3x^2\xb^2+38x\xb+8)
           \cr &\quad -23x^2\xb^2-36x\xb
          +2z \Bigl( 3(x^3+\xb^3)+x\xb(x^2+\xb^2)+
                (x+\xb) \bigl( 5x\xb-12(x+\xb)+20 \bigr)
                +2x^2\xb^2-8 \Bigr) \cr &\quad
          +8z^2 \Bigl( (x+\xb)(x+\xb-4)+2 \Bigr) \biggr)
           {x-\xb \over 2(2-x-\xb)^2(1-x)(1-\xb)(x^2-4z)(\xb^2-4z)}
           , \cr
 \tilde{b}_{24}^{v} \ &=\
                    {x-\xb \over 2(2-x-\xb)^2(1-x)(1-\xb)},
}
\eqn\rescoeffvtwotwo
$$
\vfil\eject
$$
\eqalign{
  \tilde{d}_{D=6}^{a}\ &=\
\biggl[ (x+\xb-1)^2
    - {z(2x^3-7x^2+x^2\xb+16x-11x\xb-x\xb^2+20\xb-8\xb^2-12)
            \over (1-\xb)} \cr
&\quad
+ {4z^2(14x-6x^2+x^3+16\xb-10x\xb-6\xb^2+x^2\xb-10)
            \over (x+\xb-1)(1-\xb)}
- {16z^3(2-x-\xb) \over (x+\xb-1)(1-\xb)}
  \biggr] \cr
&\quad\times
  {s \over 2(1-x)(1-\xb)(1+4z-x-\xb)}, \cr
 \tilde{d'}_{D=6}^{a} \ &=\ -\tilde{d}_{D=6}^{a}|_{x\lr\xb}, \cr
 c_{134}^{a,2} \ &=\
 \biggl[ {(\xb^2-4z)^2 \over (1-\xb)}
     -{(\xb^2-4z)(-6x-8\xb+2x\xb+1+7\xb^2) \over 2(1-\xb)}
\cr &\quad
+{-29\xb^3+11\xb^4+19x\xb+7x\xb^3-25x\xb^2-11\xb+3x+29\xb^2-2 \over
 2(1-\xb)} \cr &\quad
+{\xb(9\xb^4+8x\xb^3-27\xb^3-26x\xb^2+34\xb^2-16\xb+20x\xb+2x-2)
\over 2(\xb^2-4z)} \cr &\quad
 -{3\xb^3(1-\xb)^2(2(1-x-\xb)+\xb^2+x\xb) \over 2(\xb^2-4z)^2}
 +{(\xb^2-4z)^3 \over (x+\xb-1)(1+4z-x-\xb)(1-\xb)}
\cr &\quad
+ {(\xb^2-4z)^2(4x-5\xb^2+8\xb+x^2-6) \over
 (x+\xb-1)(1+4z-x-\xb)(1-\xb)} \cr &\quad
+ \Bigl( 3x^2\xb+2x^3-25\xb+21x\xb-23\xb^3-17x+33\xb^2+5x^2+10+8\xb^4
   -x\xb^3-4x^2\xb^2-11x\xb^2 \Bigr) \cr &\quad
\times {(\xb^2-4z) \over (x+\xb-1)(1+4z-x-\xb)(1-\xb)}
\cr &\quad
- \Bigl( 11x^2\xb^2+50x\xb-2x\xb^5+52\xb^2+46\xb^4+6+18x^2+42x\xb^3
         -67x\xb^2-5x^2\xb^4+5\xb^6 \cr &\quad
         -25\xb+4x^3\xb^2-6x^3-9x\xb^4-21\xb^5+6x^2\xb^3
         -18x-25x^2\xb-62\xb^3 \Bigr)
\cr &\quad \times {1 \over  (x+\xb-1)(1+4z-x-\xb)(1-\xb)}
\cr &\quad
+{\xb(1-\xb)(x+\xb-1-\xb^2)(2x\xb+4x-\xb^2+2\xb-4) \over  (1+4z-x-\xb)
  (\xb^2-4z)} \biggr]{s\over 8 (1-x)}  , \cr
c_{234}^{a,2}  \ &=\ -c_{134}^{a,2}|_{x\lr\xb}, \cr
\tilde{c}_{123}^a \ &=\  \biggl[
-{2x^2+2x\xb-5x-5\xb+4+\xb^2 \over 2(1-x)(1-\xb)} \cr &\quad
-{z(5x\xb^2+5x^2\xb-6x^2-10+3\xb^3+23\xb-18x\xb+x^3-16\xb^2+13x)
\over (x+\xb-1)(1-x)(1-\xb)^2}\cr &\quad
+{4z^2(2x\xb+4+x^2-3x-5\xb+\xb^2)\over (x+\xb-1)(1-x)(1-\xb)^2}
\biggr] {z s \over 1+4z-x-\xb}, \cr
\tilde{c}_{134}^{a}  \ &=\ -\tilde{c}_{123}^{a}|_{x\lr\xb}, \cr
}
\eqn\rescoeffatwoone
$$
\vfil\eject
$$
\eqalign{
\tilde{c}_{234}^{a} \ &=\
  \biggl[ -{2(x^4+\xb^4)-15(x^3+\xb^3)+(x^2+\xb^2)(7x\xb+43)
-(x+\xb)(43x\xb+52)+10x^2\xb^2+84x\xb+22 \over 2(2-x-\xb)}
\cr &\quad
+{2z\Bigl( x^4+\xb^4-9(x^3+\xb^3)+
3(x^2+\xb^2)(x\xb+10)-5(x+\xb)(8+5x\xb)
+4x^2\xb^2+58x\xb+18 \Bigr) \over (2-x-\xb)(x+\xb-1)} \cr &\quad
+{8z^2(2-x-\xb) \over x+\xb-1}
\biggr]{z(x-\xb) \over (1-x)^2(1-\xb)^2} {s \over 1+4z-x-\xb},\cr
 b_{13}^{a,2} \ &=\
 \biggl[
-{(x-\xb)(\xb^2-4z)^3 \over (1-\xb)^3} +{(\xb^2-4z)^2
(4x+x\xb^2-6\xb^3-21\xb+5+19\xb^2-2x\xb) \over (1-\xb)^3}
\cr &\quad
+{(\xb^2-4z)(-28x\xb^3+207\xb^3+62x\xb^2+6x\xb^4+197\xb+21x+15\xb^5
-277\xb^2-64x\xb-54-85\xb^4) \over (1-\xb)^3} \cr &\quad
- \Bigl( 60x-72+1293\xb^3-1055\xb^4+523\xb^5-520x\xb^3+306x\xb^4
         -98x\xb^5-956\xb^2 \cr &\quad
         +505x\xb^2-268x\xb-149\xb^6+19\xb^7+398\xb
         +14x\xb^6 \Bigr) {1\over (1-\xb)^3} \cr &\quad
- \Bigl( 12\xb^6+11x\xb^5-71\xb^5+185\xb^4-60x\xb^4
         -264\xb^3+131x\xb^3-150x\xb^2 \cr &\quad
         +220\xb^2-100\xb+88x\xb+16-16x \Bigr)
         {2-\xb \over (1-\xb)(\xb^2-4z)}
\cr &\quad
+{3\xb^2(1-\xb)(2-\xb)^3(2(1-x-\xb)+\xb^2+x\xb) \over (\xb^2-4z)^2}
\biggr]{1 \over 64(1-x) (1+z-\xb)}, \cr
  b_{24}^{a,2} \ &=\ -b_{13}^{a,2}|_{x\lr\xb}, \cr
  b_{34}^{a,2} \ &=\
 {23\xb^2+6x\xb^3-4-19x\xb^2-8\xb+7\xb^4-20\xb^3+13x\xb+4x \over
    8(1-x)(1-\xb)(\xb^2-4z)} \cr &\quad
    -{23x^2+6x^3\xb-4-19x^2\xb-8x+7x^4-20x^3+13x\xb+4\xb \over
    8(1-x)(1-\xb)(x^2-4z)} \cr &\quad
    -{3\xb^2(1-\xb)(2(1-x-\xb)+\xb^2+x\xb)\over 8(1-x)(\xb^2-4z)^2}
    +{3x^2(1-x)(2(1-x-\xb)+x^2+x\xb)\over 8(1-\xb)(x^2-4z)^2} \cr &\quad
    + \biggl( -2(x^4+\xb^4)+(x^3+\xb^3)(x\xb+10)-(x^2+\xb^2)(13x\xb+16)
              +(x+\xb)(3x^2\xb^2+38x\xb+8)
     \cr &\quad -23x^2\xb^2-36x\xb \cr &\quad
   -2z \Bigl( -2(x^4+\xb^4)+5(x^3+\xb^3)+(x^2+\xb^2)(-5x\xb+4)
    \cr &\quad +(x+\xb)(3x\xb-20)-6x^2\xb^2+24x\xb+8 \Bigr) \cr &\quad
   -8z^2 \Bigl( 3(x+\xb)(x+\xb-4)+14 \Bigr) \biggr)
{x-\xb \over 2(2-x-\xb)^2(1-x)(1-\xb)(x^2-4z)(\xb^2-4z)} \cr &\quad
+ {(x-\xb)(4(x+\xb)-1)\over 8(1-x)(1-\xb)}
           , \cr
 \tilde{b}_{24}^{a} \ &=\  \tilde{b}_{24}^{v}.
}
\eqn\rescoeffatwotwo
$$


\listrefs

\vfill\eject
\vskip0.1truein

\par\noindent
{\bf Figure Captions}
\vskip0.1truein

\par\noindent{\bf Figure 1:}
\par\noindent Sample Feynman diagram for the QCD rescattering contribution,
with $m_q\neq0$ required for a nonvanishing result.

\vskip0.1truein

\par\noindent{\bf Figure 2:}
\par\noindent Triangle diagram for QCD rescattering contribution via quark
annihilation; again $m_{q^\prime}\neq0$ is required for a nonvanishing
result.

\vskip0.1truein

\par\noindent{\bf Figure 3:}
\par\noindent
Sample diagram for electroweak rescattering contribution.

\vskip0.1truein

\par\noindent{\bf Figure 4:}
\par\noindent
Sample diagram for contribution of $\gamma\gamma$ and $\gamma Z$
intermediate states.
\vskip0.1truein

\par\noindent{\bf Figure 5:}
\par\noindent
Sample diagram for contribution from real $\gamma\gamma$ initial
state.
\vskip0.1truein

\par\noindent{\bf Figure 6:}
\par\noindent
Definition of the angles $\theta$, $\phi$ and $\thnormal$ for
$e^+e^- \to q\qb g$. The unit vector ${\bf n}$ is the (signed) normal to the
$q\bar{q}g$ plane.

\vskip0.1truein

\par\noindent {\bf Figure 7:}
\par\noindent $\costetn$ from QCD rescattering, normalized for $b$-final states
only,
as a function of the center-of-mass energy, for $\ycut = 0.04$.
The dotted line gives the
contribution of the vector component of the result.
The one loop running of $\alpha_s$ is included.

\vskip0.1truein

\par\noindent {\bf Figure 8:}
\par\noindent QCD (dotted line),
$W$-exchange (dot-dashes) and $Z$-exchange (dashes)
contributions to $\costetn$
as functions of the center-of-mass energy, for $\ycut = 0.04$.
The one loop running of $\alpha_s$ is included.
The solid line gives the sum of the three contributions.

\vskip0.1truein

\par\noindent {\bf Figure 9:}
\par\noindent QCD (solid line),
$W$-exchange (dot-dashes) and $Z$-exchange (dashes)
contributions to the signal-to-noise ratio,
divided by the square-root of the number of 2-jet events,
for $\costetn$, as functions of $\ycut$ at the $Z$.

\vskip0.1truein

\par\noindent {\bf Figure 10:}
\par\noindent $B$ boson exchange contribution
to $\costetn$ as a function of $\xi\equiv M_B^2/M_Z^2$,
with (dashed line)
and without (solid line) gluon identification,
at the $Z$, for $\alpha_B = 0.2/9$ and $\ycut = 0.04$.

\vfill\eject
\nopagenumbers

\vskip 2.0cm
\epsfxsize=2.0in
\centerline{\epsffile{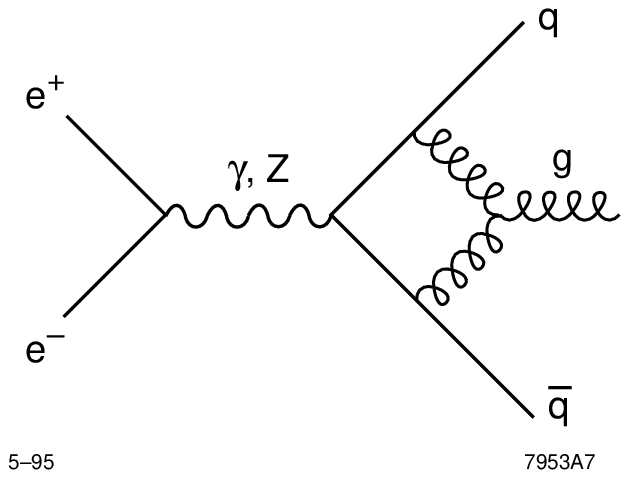}}
\vskip 0.2cm\nobreak
\centerline{\bf Fig.~1}

\vskip 2.0cm
\epsfxsize=2.0in
\centerline{\epsffile{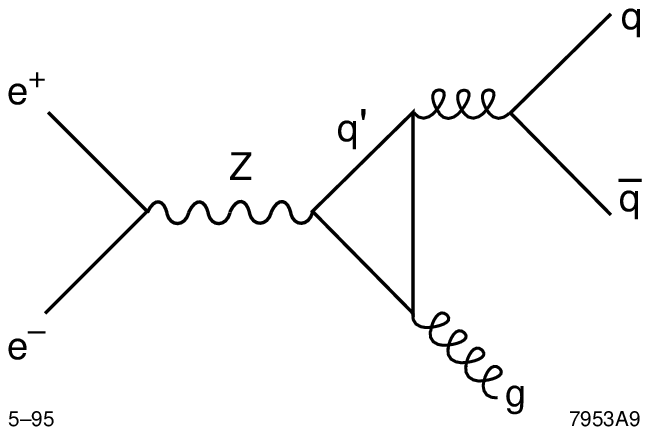}}
\vskip 0.2cm\nobreak
\centerline{\bf Fig.~2}

\vskip 2.0cm
\epsfxsize=2.0in
\centerline{\epsffile{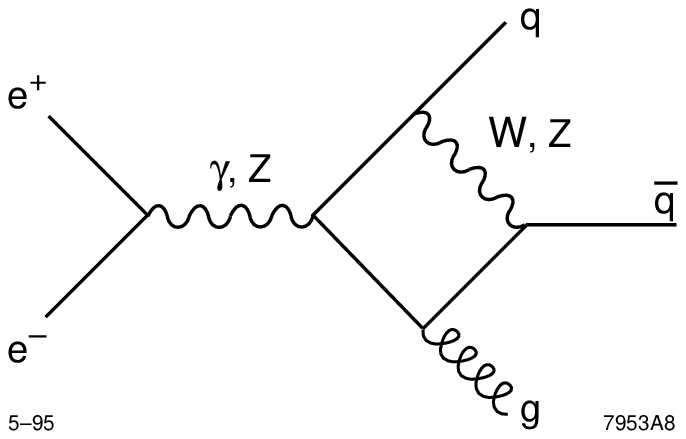}}
\vskip 0.2cm\nobreak
\centerline{\bf Fig.~3}

\vfill\eject

\vskip 2.0cm
\epsfxsize=2.0in
\centerline{\epsffile{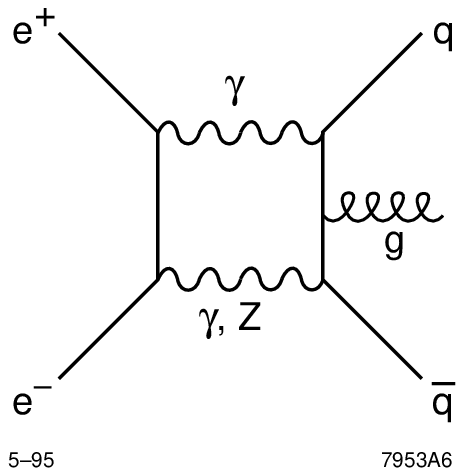}}
\vskip 0.2cm\nobreak
\centerline{\bf Fig.~4}

\vskip 2.0cm
\epsfxsize=2.0in
\centerline{\epsffile{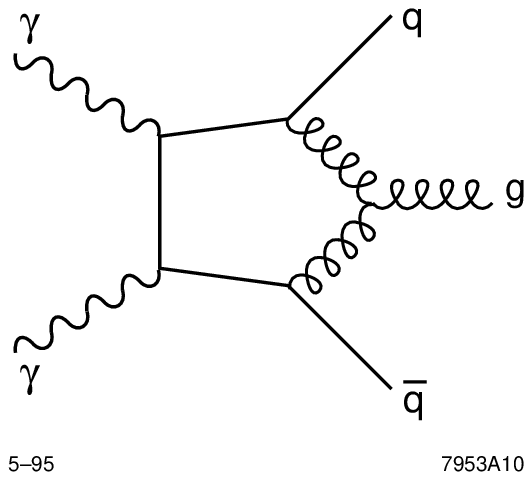}}
\vskip 0.2cm\nobreak
\centerline{\bf Fig.~5}

\vskip 2.0cm
\epsfxsize=4.0in
\centerline{\epsffile{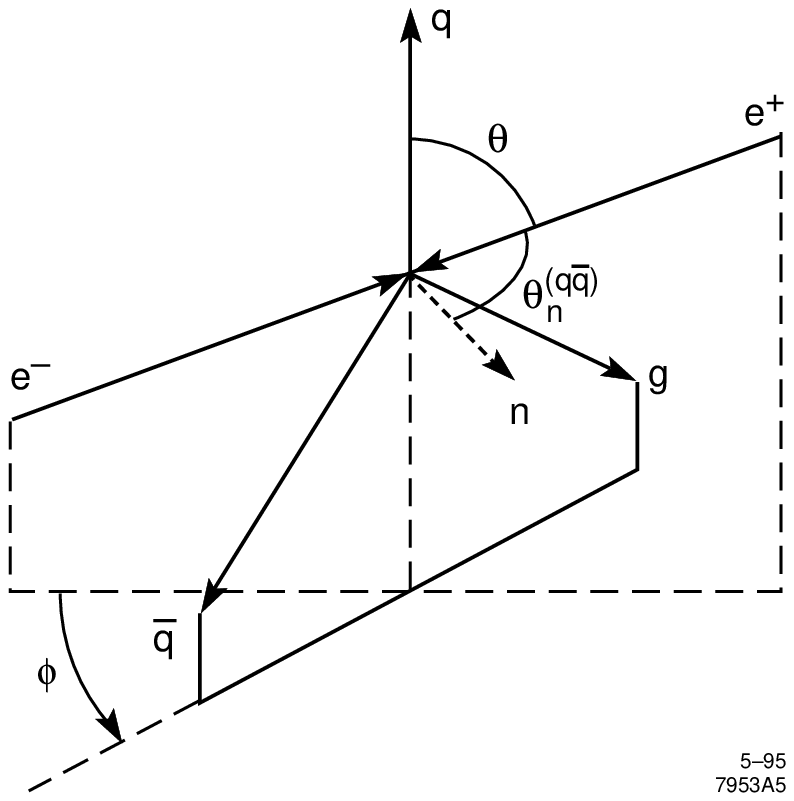}}
\vskip 0.2cm\nobreak
\centerline{\bf Fig.~6}

\vskip 2.0cm
\epsfxsize=4.0in
\centerline{\epsffile{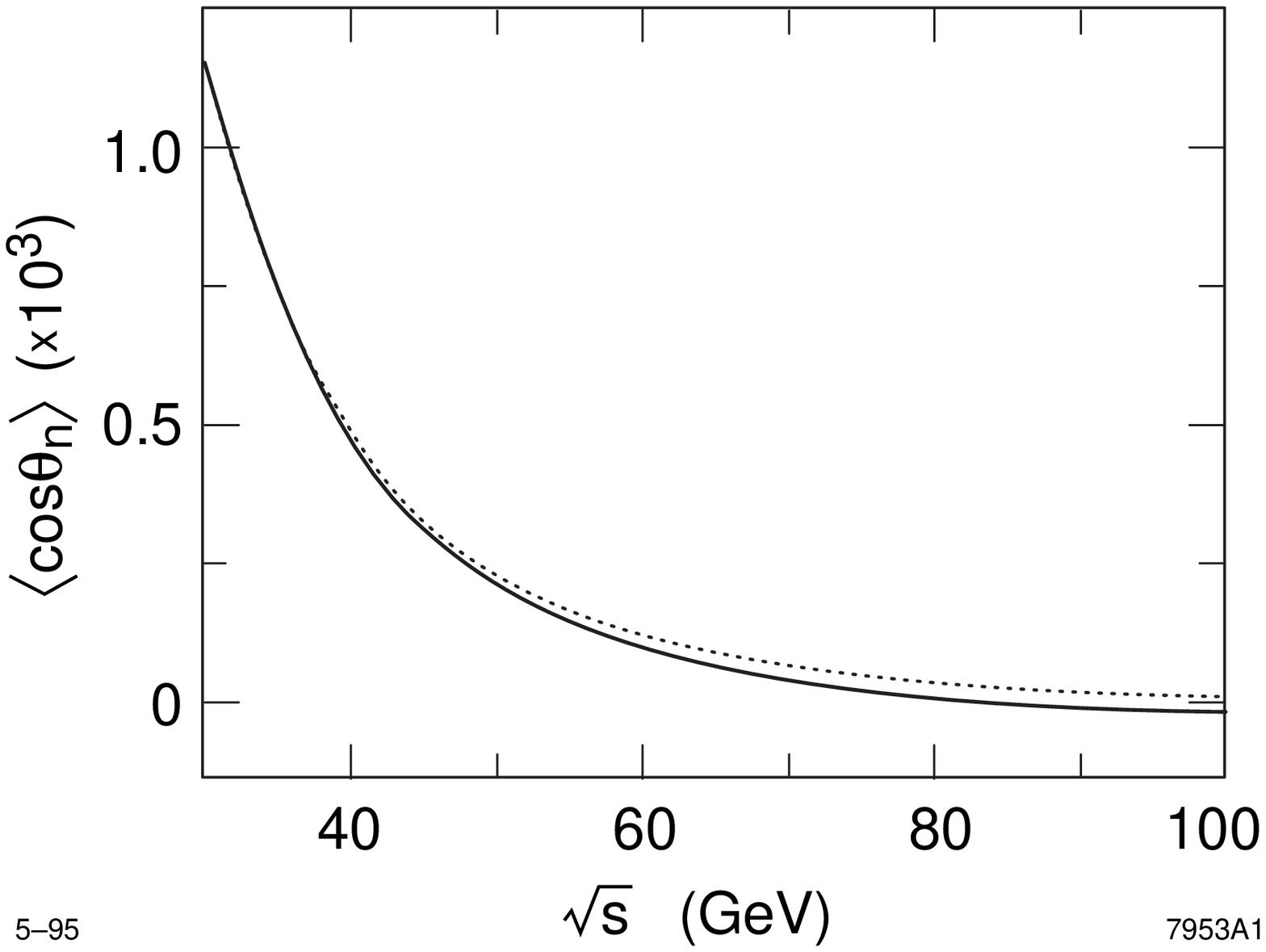}}
\vskip 0.2cm\nobreak
\centerline{\bf Fig.~7}

\vskip 2.0cm
\epsfxsize=4.0in
\centerline{\epsffile{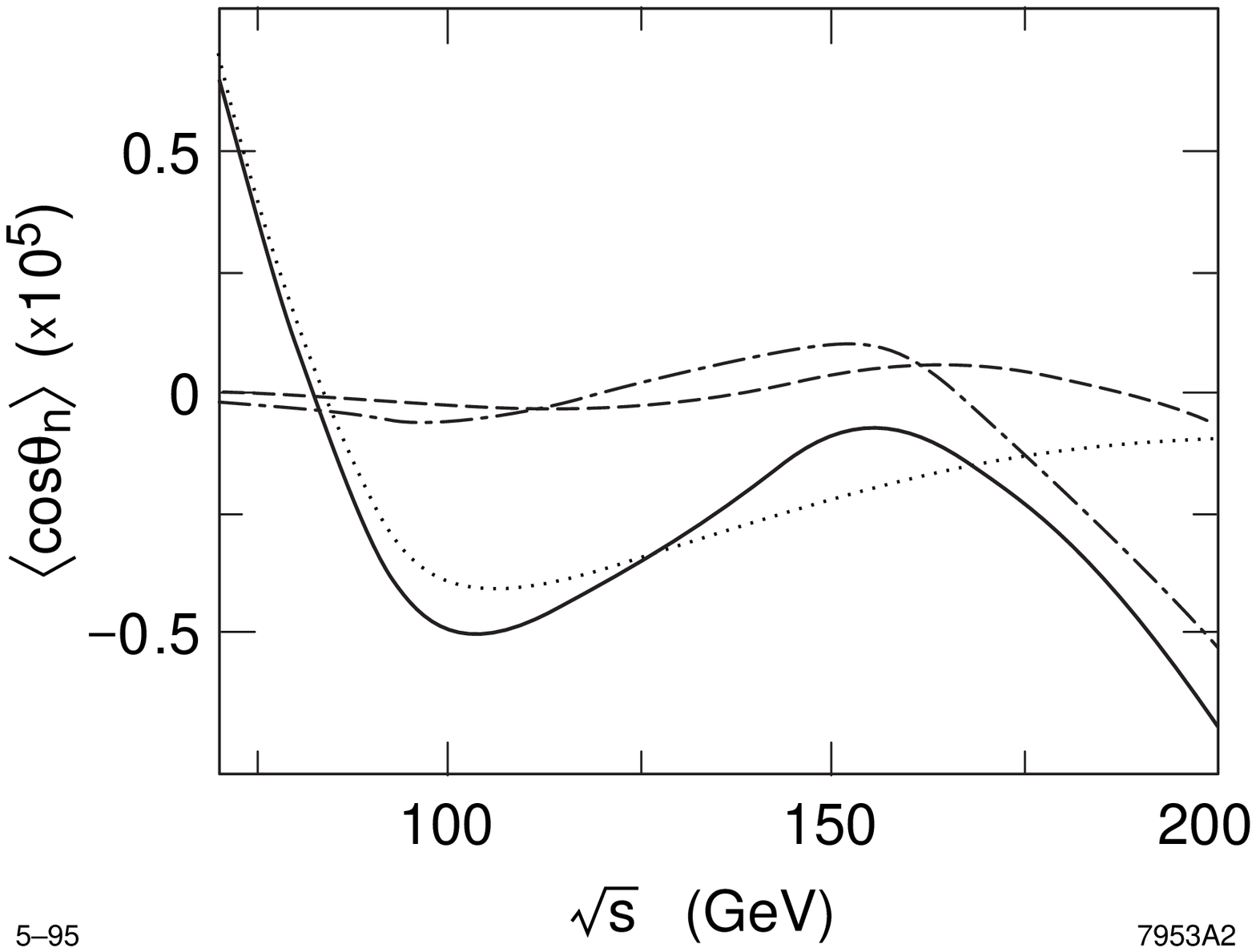}}
\vskip 0.2cm\nobreak
\centerline{\bf Fig.~8}

\vskip 2.0cm
\epsfxsize=4.0in
\centerline{\epsffile{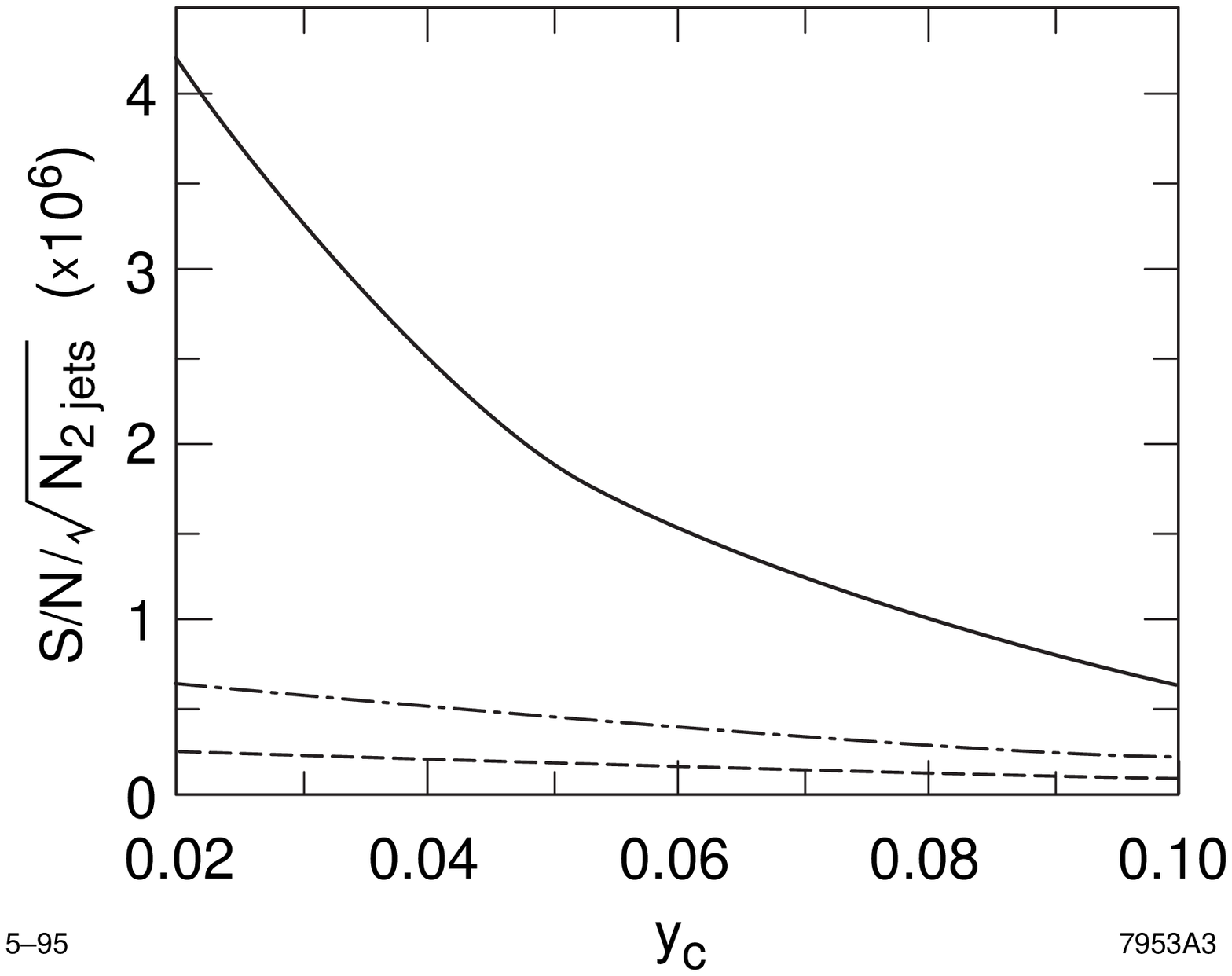}}
\vskip 0.2cm\nobreak
\centerline{\bf Fig.~9}

\vskip 2.0cm
\epsfxsize=4.0in
\centerline{\epsffile{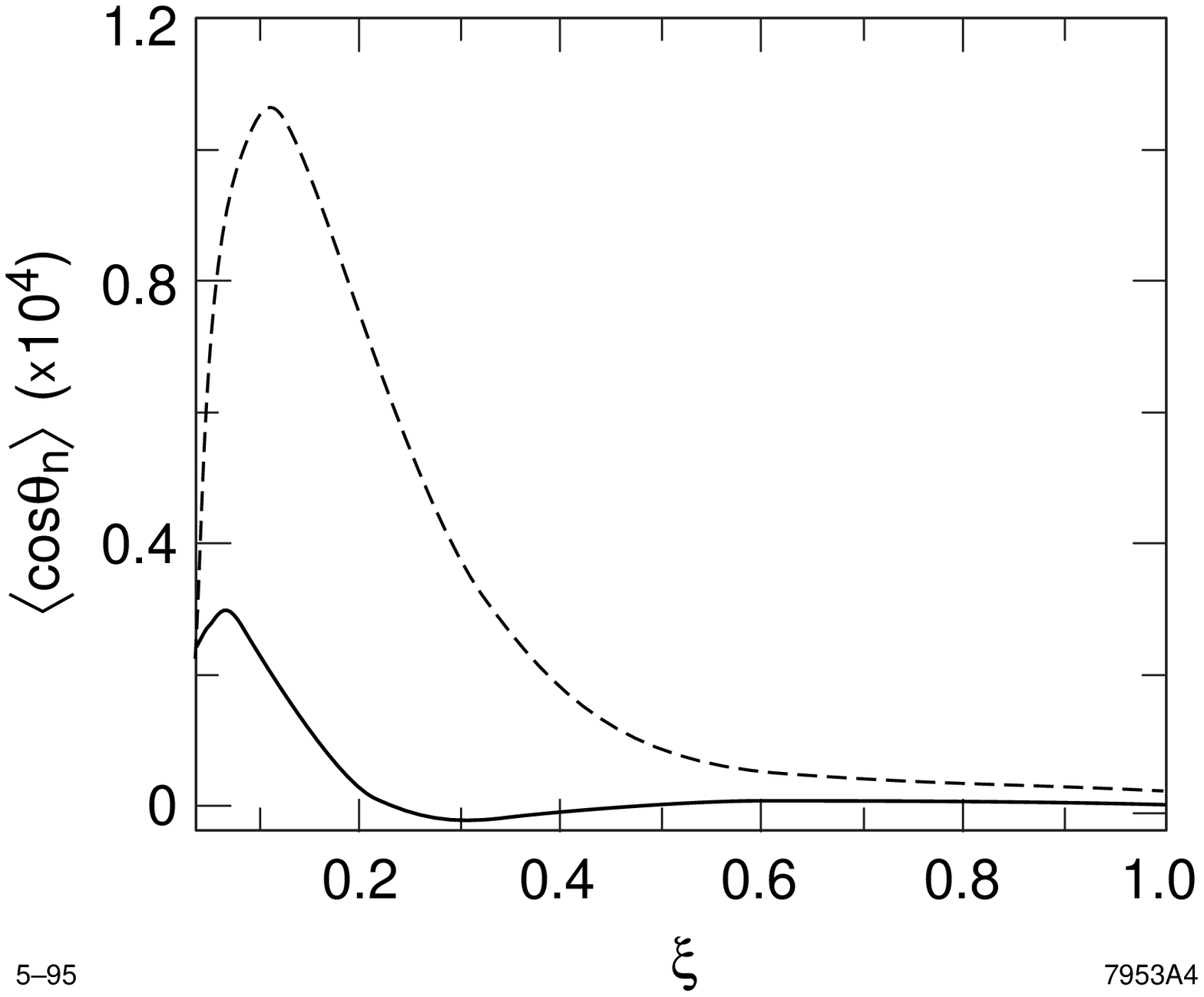}}
\vskip 0.2cm\nobreak
\centerline{\bf Fig.~10}


\end